\documentclass[
 reprint,
 superscriptaddress,
 amsmath,amssymb,
 aps,
 prx,
]{revtex4-2}

\usepackage{graphicx}% Include figure files
\usepackage{dcolumn}% Align table columns on decimal point
\usepackage{bm}% bold math
\usepackage{hyperref}% add hypertext capabilities
\usepackage{amsmath, amssymb, mathtools, isomath, nccmath}
\usepackage{siunitx}
\usepackage{xspace}
\usepackage{xcolor}
\usepackage[british]{babel}
\usepackage{sidecap}

\newcommand{\ket}[1]{\ensuremath{\lvert #1 \rangle}\xspace}%
\newcommand{\avg}[1]{\ensuremath{\langle #1 \rangle}\xspace}%
\newcommand{\abs}[1]{\ensuremath{\lvert #1 \rvert}\xspace}%

\sidecaptionvpos{figure}{h}

\begin{document}
\title{\bf{Tracking evaporative cooling of a mesoscopic atomic quantum gas in real time}} 

%\author{The ultracold E3 team}
\author{Johannes Zeiher}
\email[]{johannes.zeiher@mpq.mpg.de}
\thanks{Present address: Max-Planck-Institut f\"{u}r Quantenoptik, 85748 Garching, Germany}
\affiliation{Department of Physics, University of California, Berkeley, CA 94720, USA}

\author{Julian Wolf}
\affiliation{Department of Physics, University of California, Berkeley, CA 94720, USA}
\affiliation{Challenge Institute for Quantum Computation, University of California, Berkeley, CA 94720, USA}

\author{Joshua A. Isaacs}
\affiliation{Department of Physics, University of California, Berkeley, CA 94720, USA}
\affiliation{Challenge Institute for Quantum Computation, University of California, Berkeley, CA 94720, USA}

\author{Jonathan Kohler}
\thanks{Present address: Vector Atomic, Inc., Pleasanton, CA 94586, USA}
\affiliation{Department of Physics, University of California, Berkeley, CA 94720, USA}
\author{Dan M. Stamper-Kurn}
\affiliation{Department of Physics, University of California, Berkeley, CA 94720, USA}
\affiliation{Challenge Institute for Quantum Computation, University of California, Berkeley, CA 94720, USA}
\affiliation{Materials Science Department, Lawrence Berkeley National Laboratory, Berkeley, CA 94720, USA}

\date{\today}

%%%%%%%%%%%%%%%%%%%%%%%%%%%%%%%%%%%%%%%%%%%%%%%%%%%%%%%%%%%%%%%%%%%%%%%
%                        Summary paragraph                            %
%%%%%%%%%%%%%%%%%%%%%%%%%%%%%%%%%%%%%%%%%%%%%%%%%%%%%%%%%%%%%%%%%%%%%%%

\begin{abstract}
The fluctuations in thermodynamic and transport properties in many-body systems gain importance as the number of constituent particles is reduced.
Ultracold atomic gases provide a clean setting for the study of mesoscopic systems; however, the detection of temporal fluctuations is hindered by the typically destructive detection, precluding repeated precise measurements on the same sample.
Here, we overcome this hindrance by utilizing the enhanced light--matter coupling in an optical cavity to perform a minimally invasive continuous measurement and track the time evolution of the atom number in a quasi two-dimensional atomic gas during evaporation from a tilted trapping potential.
We demonstrate sufficient measurement precision to detect atom number fluctuations well below the level set by Poissonian statistics. Furthermore, we characterize the non-linearity of the evaporation process and the inherent fluctuations of the transport of atoms out of the trapping volume through two-time correlations of the atom number.
Our results establish coupled atom--cavity systems as a novel testbed for observing thermodynamics and transport phenomena in mesosopic cold atomic gases and, generally, pave the way for measuring multi-time correlation functions of ultracold quantum gases. 
\end{abstract}

\maketitle

%%%%%%%%%%%%%%%%%%%%%%%%%%%%%%%%%%%%%%%%%%%%%%%%%%%%%%%%%%%%%%%%%%%%%%%
%                        General Introduction                         %
%%%%%%%%%%%%%%%%%%%%%%%%%%%%%%%%%%%%%%%%%%%%%%%%%%%%%%%%%%%%%%%%%%%%%%%
\section{Introduction}
Tracking out-of-equilibrium dynamical processes and their fluctuations in mesoscopic systems is central to thermodynamics at intermediate scales~\cite{Campisi2011,Campisi2010} and transport in solid state systems~\cite{Blanter2000}. 
For example, current fluctuations in mesoscopic electronic devices reveal the charge quantization of elementary or emergent particles, shedding light on the underlying microscopic physics~\cite{Schottky1918,Gustavsson2006}.         
Advanced experimental control and precise measurements make ultracold atomic gases an ideal testbed for studying transport phenomena with solid-state analogs and beyond~\cite{Brantut2012,Brantut2013,Chien2015}. Furthermore, the achievable system sizes, ranging from single to millions of atoms in different setups, naturally provide access to explore the mesoscopic domain with ensembles of cold atoms.\\
However, solid-state and ultracold-atom mesoscopic systems differ in their fragility against measurement. Solid-state devices are coupled to large thermal reservoirs, which rapidly dissipate the backaction of measurement, and are refreshed with large particle reservoirs. In contrast, ultracold-atom systems are well isolated from thermal environments. Technical and backaction disturbance from measurement, such as from optical force fluctuations caused by light scattering, is absorbed within the mesoscopic system itself and can change the properties of the system significantly. Thus, the measured fluctuations within a mesoscopic cold-atom system can be strongly altered by continuous or stroboscopic measurements performed on the system.
Accessing real-time information in such systems therefore requires strategies to maximize the extracted information for a given heating rate associated with the measurement.\\
The enhanced atom--light interaction in high-finesse optical cavities~\cite{Tanji-Suzuki2011} provides a means for performing minimally invasive, extremely sensitive measurements on atomic gases. Demonstrations span from recording transient signals of single or few atoms passing through an optical cavity~\cite{Hood2000,Pinkse2000} to measurements on static and dynamically evolving mesoscopic trapped atomic ensembles~\cite{Brahms2011,Zhang2012,Chen2014,Norcia2016,Kohler2017,Roux2021}, or the probing of dynamical evolution of novel states of matter realized in the cavity~\cite{Leonard2017c,Leonard2017b,Kroeze2019}.
Cavity-enhanced dispersive atom number measurements have been proposed as a non-invasive probe for dynamical and transport phenomena in mesoscopic samples~\cite{Krinner2017,Uchino2018,Yang2018}. The achievable precision surpasses that of stroboscopic free-space dispersive measurements, which have been performed on macroscopic atomic gases with greater capacity than mesoscopic gases to absorb backaction disturbance~\cite{Sawyer2012,Gajdacz2013,Kristensen2019,Christensen2021}.
Cavity-enhanced atom-number readout with single-atom precision has been demonstrated, but was accompanied by strong disturbance that precluded repeated measurement on the gas~\cite{Zhang2012}.\\
In this work, we employ cavity-enhanced measurements to continuously track the non-equilibrium process of evaporative cooling~\cite{Davis1995b, Luiten1996a,Ketterle1996a, Wu1996, OHara2001a} of a mesoscopic sample for long evolution times, which allows us to probe temporal correlations at all times under observation.
Evaporative cooling occurs in a gas of temperature $T$ and atom number $N$ when collisions drive atoms to energies above the finite trap depth $U$, whereupon these atoms escape the trap, reducing the number of atoms remaining as well as their temperature. 
%%%%%%%%%%%%%%%%%%%%%%%%%%%%%%%%%%%%%%%%%%%%%%%%%%%%%%%%%%%%%%%%%%%%%%%
%                       	  Figure 1  include                       %
%%%%%%%%%%%%%%%%%%%%%%%%%%%%%%%%%%%%%%%%%%%%%%%%%%%%%%%%%%%%%%%%%%%%%%%
\begin{figure*}[t!]
  %\centering
  \includegraphics[width=\textwidth]{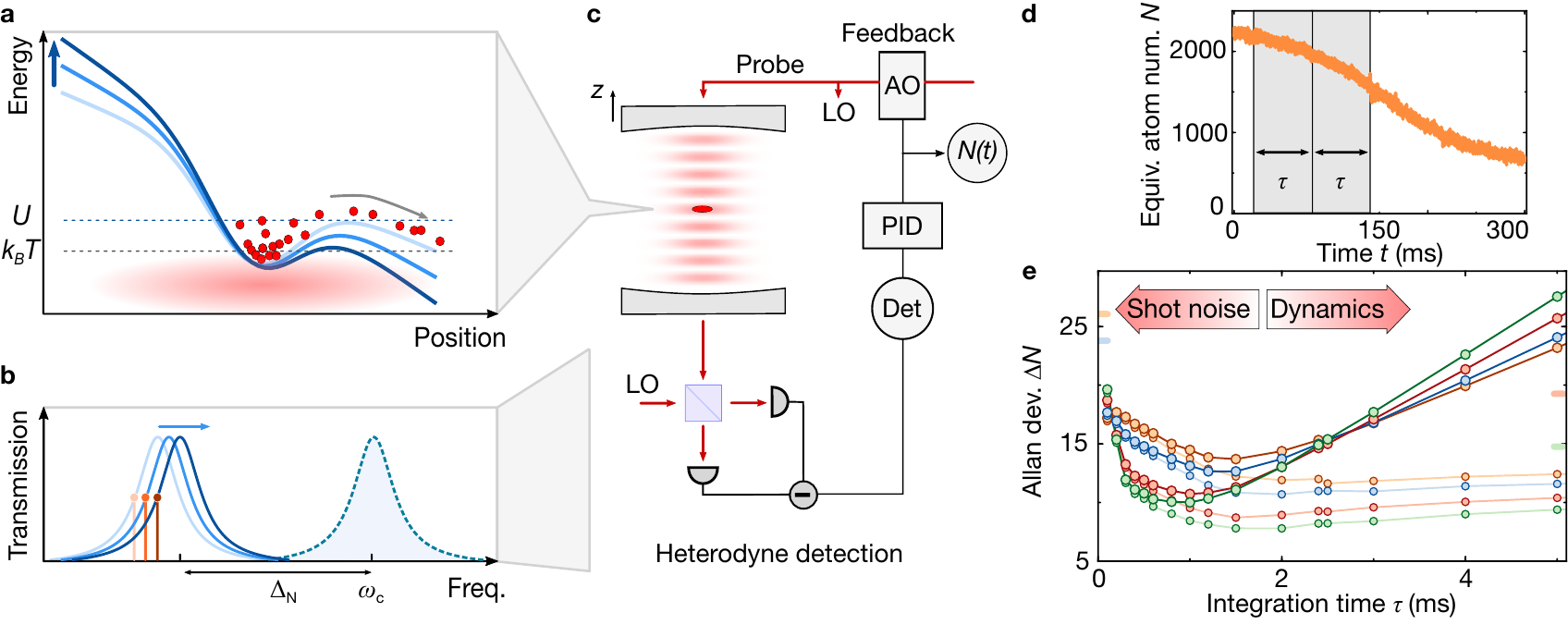}%
  \caption{ \label{fig:1}
  \textbf{Experimental setup and measurement imprecision.}
  \textbf{a} Schematic of the tilted evaporation of atoms (red) with temperature $T$ in the trap potential (blue) with  dynamically lowered depth $U$. Atoms with energies exceeding the trap depth are spilled. \textbf{b} The atoms dispersively couple to the cavity, resulting in an atom-number-dependent shift $\Delta_N$ of the transmission line shape (blue solid traces) compared to an empty cavity (blue dashed line). Tracking the side of fringe (marked by red dot and line) provides a dynamical measurement of atom number. \textbf{c} The cavity resonance is tracked using a feedback loop involving the cavity-coupled atomic cloud (dark red), located at the intensity maximum of a probe beam in the cavity (light red). The transmitted intensity of probe light is detected using a heterodyne receiver with local oscillator (LO) and a radiofrequency power detector (Det). This power is kept constant using a feedback loop (PID) adjusting the frequency of probe and LO through an acousto-optic modulator (AO). The atom number is derived from an in-loop measurement of the voltage used to adjust a voltage controlled oscillator driving the AO. \textbf{d} A single unfiltered trace of equivalent atom number $N$ vs.~time with the filter procedure indicated (gray shaded areas). \textbf{e} The Allan deviation $\Delta N$ is dominated by photonic shot noise for small integration times $\tau$ and by dynamics of the atom number for large $\tau$. A low pass filter of the traces with optimal integration time $\tilde{\tau}$ minimizes the noise associated with both effects. The solid lines represent guides to the eye. Larger photon number $n$ (orange: $n=1.9(1)$, blue: $n=3.2(1)$, red: $n=6.0(1)$ and green: $n=9.7(1)$) results in reduced shot noise, but also faster loss of the atoms and therefore increased imprecision at longer integration times. The colored ticks mark the noise level set by Poissonian statistics for our lowest measured mean atom number for each $n$. The light symbols show a reduced Allan deviation, where the average dynamics has been subtracted from each trace before calculating $\Delta N$.}
\end{figure*}  
The ensuing dynamics depend on dimensionality, atom number, and temperature of the gas, all features also at the heart of transport phenomena studied with destructive measurements~\cite{Krinner2017}.
A simple model captures the interplay between temperature and atom number in an evaporatively cooled atomic ensemble~\cite{Ketterle1996a}: To evaporate from the trap, atoms have to be collisionally transferred to the high-energy tail of the Maxwell--Boltzmann distribution, such that the average evaporation rate, $\dot{N}\propto -\eta e^{-\eta}$, depends exponentially on $\eta=U/k_BT$. This implies that samples with initially higher temperature evaporate atoms more quickly than samples with lower temperature. Moreover, due to the density-dependent thermalization rate of an evaporatively cooled gas~\cite{Luiten1996a,Ketterle1996a} and the presence of three-body collisions~\cite{Whitlock2010}, the evaporation dynamics can be expected to be non-linear in atom number. In the course of evaporative cooling, thermodynamic properties of the trapped atomic gas also undergo stochastic fluctuations. For example, a linear single-particle loss process results in a binomial partition of the gas between trapped and untrapped populations, with fluctuations described by the binomial distribution. Non-linearity associated with few-body loss processes, and also the non-linear dependence of evaporative cooling on instantaneous atom number and temperature, can be expected to modify these fluctuations, which are particularly pronounced in mesoscopic systems.\\
Here, we observe the non-equilibrium dynamics of an ultracold quantum gas during forced evaporation in a tilted trap potential by collecting real-time traces of the atom number dynamics (see Fig.~\ref{fig:1}a). We utilize the enhanced optical cross section of atoms coupled to a single mode of a high-finesse optical cavity to perform minimally invasive high-precision atom number measurements. In particular, we reveal two distinct dynamical regimes during evaporation: first, a super-linear regime driven by temperature variations early in the evaporation process, and, second, a sub-linear regime when these variations are damped away. Furthermore, we directly observe the temporal growth of stochastic fluctuations inherent in the evaporative cooling process itself.
%%%%%%%%%%%%%%%%%%%%%%%%%%%%%%%%%%%%%%%%%%%%%%%%%%%%%%%%%%%%%%%%%%%%%%%
%                       	  Figure 1 description                    %
%%%%%%%%%%%%%%%%%%%%%%%%%%%%%%%%%%%%%%%%%%%%%%%%%%%%%%%%%%%%%%%%%%%%%%%  
\section{Dispersive atom number readout}
%%%%%%%%%%%%%%%%%%%%%%%%%%%%%%%%%%%%%%%%%%%%%%%%%%%%%%%%%%%%%%%%%%%%%%%
%                       	  Figure 2 include                        %
%%%%%%%%%%%%%%%%%%%%%%%%%%%%%%%%%%%%%%%%%%%%%%%%%%%%%%%%%%%%%%%%%%%%%%%
\begin{figure*}[t!]
  \centering
  \includegraphics[width=\textwidth]{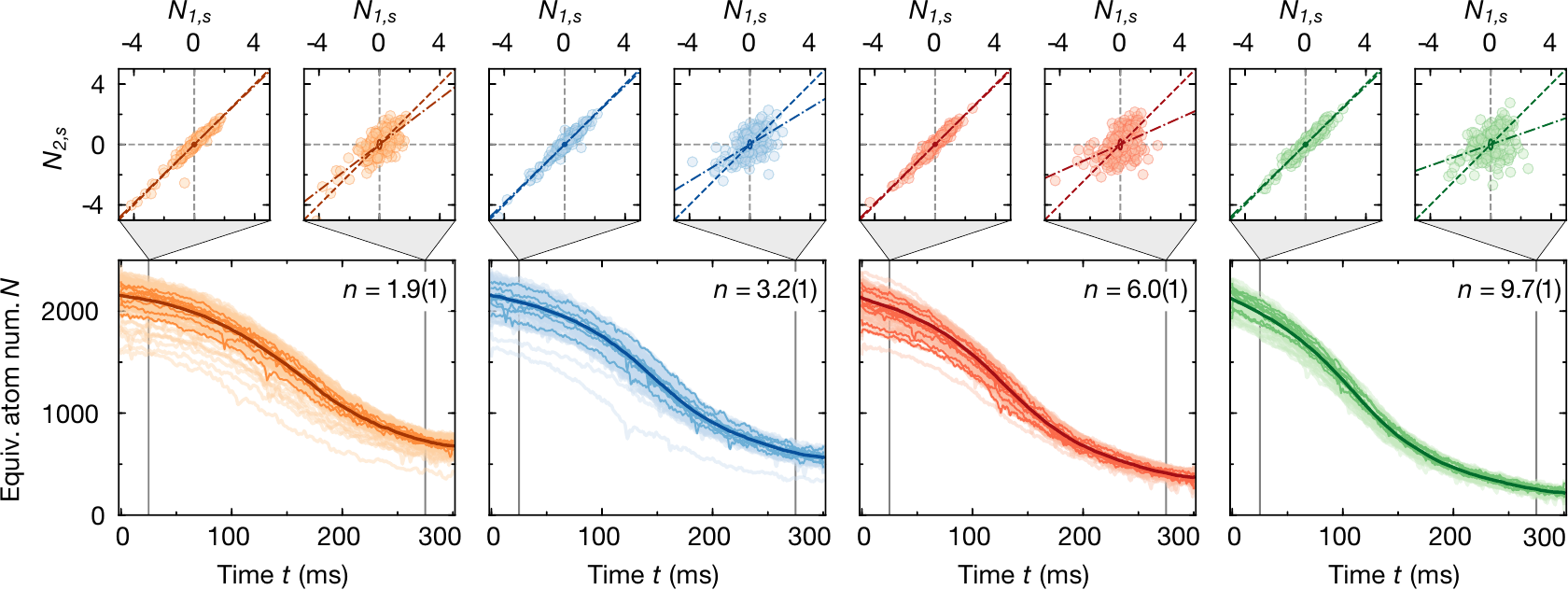}%
  \caption{ \label{fig:2}
  \textbf{Two-time correlations of atom numbers from real-time trajectories.}
  \textbf{a} Traces of equivalent atom number $N$ vs.~observation time $t$, calculated from the number-dependent cavity shift, for varying intracavity photon number $n$ (indicated in the upper right corner). The bright curves represent individual runs of the experiment (approximately $150$ per panel). Selected traces are shown in slightly darker color to illustrate their stochastic character. The dark curves are the mean over all traces. The traces were filtered using their corresponding optimal integration time; see Fig.~\ref{fig:1}e. The vertical lines mark the two times $t_2=25\,$ms ($t_2=275\,$ms) used to produce the scatter plots shown above the time traces. They illustrate the standardized atom number $N_{2,s}$ measured at the earlier (later) $t_2$ vs.~$N_{1,s}$ measured at $t_1=5\,$ms. The diagonal (dashed line) and the slope extracted from a linear ordinary least squares fit to the data (dash-dotted line) are indicated in all scatter plots. The corresponding measurement noise is indicated by the ellipse in the center.}
\end{figure*}
The principle of continuous dynamical dispersive atom number readout of our atomic cloud is illustrated in Fig.~\ref{fig:1}b and c.
We consider a cloud of atoms that is localized at the intensity maximum of the standing wave formed by a probe beam coupled into the TEM$_{00}$ mode of a Fabry--Perot optical cavity. The cavity resonance is at a frequency $\omega_c$ in the absence of atoms, which are coupled to the cavity with a single-atom vacuum Rabi coupling $g$ on an optical transition with frequency $\omega_a$, linewidth $\Gamma$ and atom-cavity detuning $\Delta_{ca} = \omega_c-\omega_a$. In the dispersive limit $\abs{\Delta_{ca}}\gg \sqrt{N}\,g\gg\Gamma$, the presence of $N$ atoms causes a frequency shift of the light-like mode of the coupled atom-cavity system by an amount proportional to the atom number~\cite{Brahms2011, Zhang2012,Chen2014},
\begin{align}
\label{eq:Delta}
\Delta_N = N\frac{g^2}{\Delta_{ca}}.
\end{align}
We detect this dispersive shift by feedback-stabilizing the frequency of the weak cavity probe to the side of fringe of the cavity transmission, realizing a continuous quantum non-demolition measurement of atom number~\cite{Chen2014}. During the measurement, the probe power in the cavity is held constant at a level characterized by the intracavity photon number $n = P/(2\hbar\omega_c\kappa\epsilon_c)$, where $\kappa$ denotes the cavity half linewidth, $P$ is the power transmitted through the cavity and $\epsilon_c$ quantifies the photon extraction efficiency. The frequency shift of the cavity resonance is extracted through the feedback signal; see Fig.~\ref{fig:1}b and Fig.~\ref{fig:S1}.
We use knowledge of $g$~\cite{Brahms2011,Kohler2017,Kohler2018} and $\Delta_{ca}$ to calculate the dispersive cavity shift due to a single atom, $\Delta_1$, and estimate the instantaneous ``equivalent atom number'' $N(t)=\Delta_{N(t)}/\Delta_1$, see Fig.~\ref{fig:1}d, where ``equivalent'' reflects a slight reduction of the vacuum Rabi coupling in our experiment (see Appendix A) and is implied if not stated explicitly otherwise.\\
We monitor the evolution of the intracavity gas during evaporation dynamics initiated by an applied magnetic field gradient~\cite{Hung2008a}; see Fig.~\ref{fig:1}a.
In the following, we discuss different quantities used to extract dynamics, measurement noise and intrinsic fluctuations from our data.
Commonly adopted measures to quantify the imprecision in the real-time measurement for different integration times are the Allan deviation $\Delta N(\tau) =(\langle \left(N_\tau(t_{i+1})- N_\tau(t_i)\right)^2\rangle/2)^{1/2}$~\cite{Brahms2011,Zhang2012} or the corresponding Allan variance $\Delta N^2$. Here, the trace $N_\tau(t_i)$ is obtained by low-pass filtering the full trace $N(t)$ with an integration time $\tau$ and then resampling at discrete times $t_i$ separated by time intervals of length $\tau$. The angle brackets denote the average over all $i$. Photon shot noise limits the measurement precision of the dispersive cavity shift for short integration times; see Fig.~\ref{fig:1}e. 
For integration times above approximately $1\,$ms, the dynamics of the evaporation process start dominating the Allan deviation.
Choosing such a long integration time leads to a loss of information about the dynamical system under observation.
As a consequence, the dynamics set an upper bound on the achievable integration times and therefore suppression of photonic shot noise in the measurement; see Appendix D. Despite this, the minimal imprecision and therefore the measurement noise of our cavity-assisted detection is well below the level set by Poissonian fluctuations of size $\sqrt{N}$ for $N$ atoms for all measured traces and all times presented in the following; see Fig.~\ref{fig:1}e.
The effect of dynamics on the optimal integration time can be mitigated by subtracting the average dynamics from each trace. The reduced Allan deviation calculated from these subtracted traces lacks the strong increase towards longer integration times, such that, in this case, the optimal integration times are typically longer; see Fig.~\ref{fig:1}e. As described in detail below, a generalization of this method with an adjusted time separation between the two integration intervals detects the intrinsic stochastic fluctuations inherent in a dynamically evolving system.\\   
%%%%%%%%%%%%%%%%%%%%%%%%%%%%%%%%%%%%%%%%%%%%%%%%%%%%%%%%%%%%%%%%%%%%%%%
%                        Physical system                              %
%%%%%%%%%%%%%%%%%%%%%%%%%%%%%%%%%%%%%%%%%%%%%%%%%%%%%%%%%%%%%%%%%%%%%%%
We started our experiment with an atomic cloud of about $2200$ $^\mathrm{87}$Rb atoms with a mean temperature of approximately $T_0 = 2.6\,\mu$K. The gas was prepared in its hyperfine state $\ket{F,\,m_F}=\ket{2,\,2}$ and was trapped predominantly in a single well of a far-detuned optical lattice potential with an initial depth $U_0/k_B = 31(1)\,\mu$K in an optical cavity~\cite{Purdy2010a,Kohler2017,Kohler2018}. The lattice provided strong confinement in the $z$-direction with a trapping frequency of $\omega_z/2\pi = 91(2)\,$kHz, putting the gas in the quasi two-dimensional regime $\hbar\omega_z>k_BT_0$. At these parameters, we estimate the phase-space density to be $0.3$, close to the quantum degenerate regime.
Crucially, accurate positioning of the atomic cloud along the cavity axis at the peak of the probe standing wave maximized the atom-cavity coupling, rendered it nearly identical for all atoms, and minimized optomechanical backaction heating~\cite{Purdy2010a,Brahms2012}.
The cavity length was stabilized such that its resonance frequency was kept at a near-constant red detuning $\Delta_{ca}/2\pi\approx-42\,$GHz with respect to the $D2$ line of $^\mathrm{87}$Rb, for which the atomic resonance linewidth is $\Gamma/2\pi \approx 6\,$MHz. The detuning of the cavity probe from atomic resonance together with the vacuum Rabi coupling $g/2\pi=13.1\,$MHz led to a maximal cavity shift $\abs{\Delta_{1}/2\pi}\approx4\,$kHz per atom.
The cavity probe was maintained at a constant detuning of $\delta_{pc}/2\pi=\kappa/2\pi=1.8\,$MHz from the atom-shifted cavity resonance frequency. The frequency lock of the probe to the cavity was realized by stabilizing the radiofrequency power output from a heterodyne receiver monitoring the probe transmission through the cavity; see Fig.~\ref{fig:1}c and Appendix A. 
The large single-atom cooperativity $C = g^2/\kappa\Gamma = 15.9$ and consequently low cavity probe powers of a few picowatts in our experiment allowed us to minimize the probe-induced off-resonant scattering and associated heating rate. This sets the cavity-based measurement apart from free-space methods and was key for achieving the long measurement times at the imprecision required for performing atom counting in mesoscopic system; see Appendix D.\\
%%%%%%%%%%%%%%%%%%%%%%%%%%%%%%%%%%%%%%%%%%%%%%%%%%%%%%%%%%%%%%%%%%%%%%%
%                       	  Figure 2                                %
%%%%%%%%%%%%%%%%%%%%%%%%%%%%%%%%%%%%%%%%%%%%%%%%%%%%%%%%%%%%%%%%%%%%%%%
In a first set of experiments, we tracked the equivalent atom number for different intracavity photon numbers $n$, while slowly lowering the trap potential $U/k_B$ from $31\,\mu$K to approximately $8\,\mu$K by ramping up a magnetic field gradient within $330\,$ms; see Fig.~\ref{fig:S2} and Appendix A.
The resulting ensemble of atom number traces is shown in Fig.~\ref{fig:2}, with every individual trace representing a new run of the experiment. The traces taken together form a statistical ensemble that encompasses both the variation in evaporation trajectories with different initial atom number and temperature, and also the fluctuations in atom number generated by evaporation dynamics in individual trajectories. In order to minimize the effect of noise, each trace was filtered with a bandwidth corresponding to the optimal integration time $\tilde{\tau}$ extracted from the Allan deviation shown in Fig.~\ref{fig:1}c. For clarity we suppress the subscript and write $N(t)\equiv N_{\tilde{\tau}}(t)$ in the following.
Our measurements of $N(t)$ show a clear trend to lower final atom numbers as the intracavity photon number is increased. This trend reflects the larger measurement-induced heating at increasing probe power, which leads to an increase in the number of atoms ejected from the trap during evaporation.\\
In general, decorrelation of two measured atom numbers $N(t_1)\equiv N_1$ and $N(t_2)\equiv N_2$ at two points in time, indicated by $t_1$ and $t_2$, arises from three sources: technical noise, measurement noise and stochastic noise due to the evaporation process itself. We extract the correlations from the slope of the standardized atom numbers $N_{2,s}$ vs.~$N_{1,s}$, which equals the Pearson correlation coefficient $\rho_{12} = \mathrm{cov}(N_1,N_2)/\sigma_1\sigma_2$; see Appendix C. The standardized atom numbers are defined as $N_{1(2),s} = (N_{1(2)}-\avg{N_{1(2)}})/\sigma_{1(2)}$, where $\avg{N_1}$ ($\avg{N_2}$) and $\sigma_{1}$ ($\sigma_2$) are the mean and the standard deviation of the non-standardized atom number distribution measured at time $t_1$ ($t_2$). In our experiment, we observe dominant linear correlations of around $\rho_{12}=97\%$ at all intracavity photon numbers for two measurements closely spaced in time; see Fig.~\ref{fig:2} upper panel. 
The strong observed correlation indicates a small influence of all noise sources at these early times. In particular, it confirms that our measurement noise is small, consistent with our previous analysis of the imprecision. For a time $t_2=275\,$ms, later on the evaporation trajectory, the fluctuations accrued over the process of evaporation are much more prominent. However, there is still some degree of linear correlation present. The reduction of the correlation from $\rho_{12}=75.9(1)\%$ at the smallest intracavity photon number $n=1.9(1)$ to $35.0(1)\%$ at the largest intracavity photon number $n=9.7(1)$ indicates a larger impact of stochastic noise coupled into the system at larger lost fraction of atoms.
Extracting the final temperature of the evaporating ensemble from a time of flight absorption image after our real-time measurement, we estimate the temperature to drop to less than $1.5(1)\,\mu$K for $n=1.9(1)$, which implies a phase-space density increase of at least $13\%$, but possibly significantly more; see Fig.~\ref{fig:S4} and Appendix A.  
%%%%%%%%%%%%%%%%%%%%%%%%%%%%%%%%%%%%%%%%%%%%%%%%%%%%%%%%%%%%%%%%%%%%%%%
%                       	  Figure 3                                %
%%%%%%%%%%%%%%%%%%%%%%%%%%%%%%%%%%%%%%%%%%%%%%%%%%%%%%%%%%%%%%%%%%%%%%%
%%%%%%%%%%%%%%%%%%%%%%%%%%%%%%%%%%%%%%%%%%%%%%%%%%%%%%%%%%%%%%%%%%%%%%%
%                       	  Figure 3 include                        %
%%%%%%%%%%%%%%%%%%%%%%%%%%%%%%%%%%%%%%%%%%%%%%%%%%%%%%%%%%%%%%%%%%%%%%%
\begin{figure}[t!]
  \centering
  \includegraphics{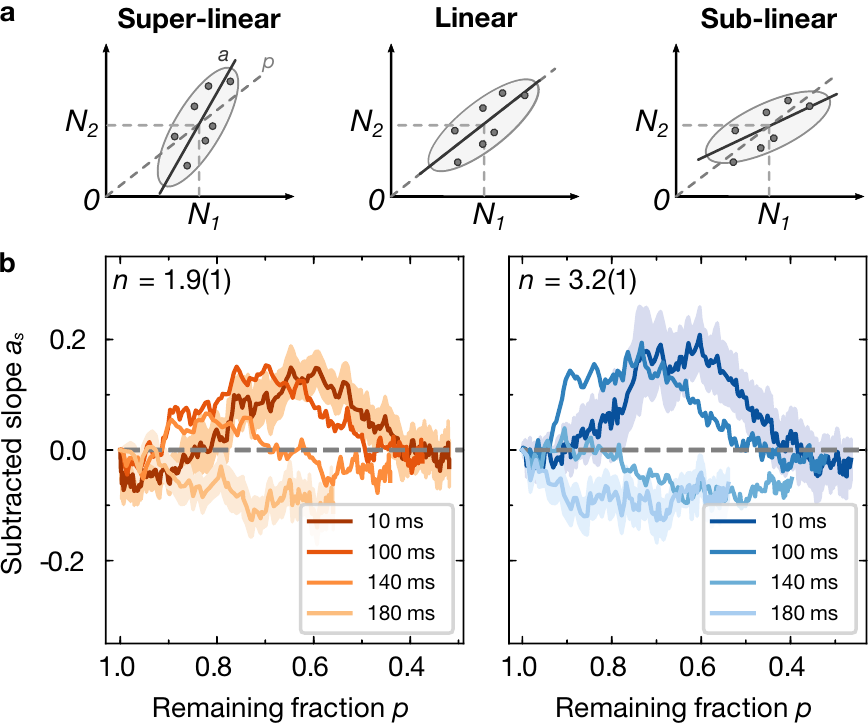}
  \caption{ \label{fig:3}
  \textbf{Non-linearity of evaporation process.}
\textbf{a} Schematic illustrating the non-linearity measure. Super- (sub-) linearity is equivalent to a slope $a$ of the scatter of $N_2$ vs.~$N_1$ that is larger (smaller) than the remaining fraction of atoms $p$. For a linear process, $a=p$. 
\textbf{b} Subtracted slope $a_s=a-p$ for different initial times $t_1$ (dark to bright color, indicated in legend) used to quantify the non-linearity of the evaporation process vs.~remaining fraction of atoms $p$. The shaded region shows the standard deviation calculated by bootstrapping for the smallest and largest initial time.}
\end{figure}
%%%%%%%%%%%%%%%%%%%%%%%%%%%%%%%%%%%%%%%%%%%%%%%%%%%%%%%%%%%%%%%%%%%%%%%
%                       	  Figure 4 include                        %
%%%%%%%%%%%%%%%%%%%%%%%%%%%%%%%%%%%%%%%%%%%%%%%%%%%%%%%%%%%%%%%%%%%%%%%
\begin{figure*}[t!]
  \centering
  \includegraphics{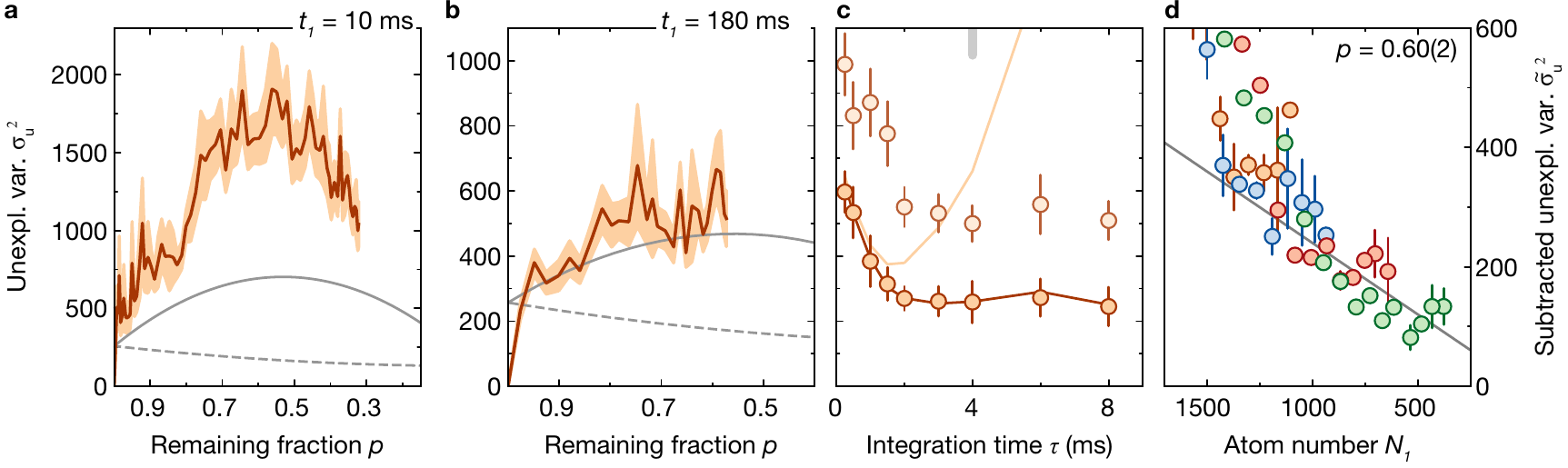}%
  \caption{\label{fig:4}
  \textbf{Mesoscopic fluctuations of the evaporation process.}
  \textbf{a} Unexplained variance $\sigma_u^2$ vs.~remaining fraction of atoms $p$ for intracavity photon number $n=1.9(1)$ at initial time $t_1=10\,$ms and \textbf{b} at $t_1=180\,$ms, with the standard deviation calculated by bootstrapping indicated by the shaded region. The solid gray lines show the model prediction including Poissonian stochastic noise and measurement noise, the dashed lines show the prediction without the contribution of Poissonian stochastic noise. \textbf{c} Unexplained variance $\sigma_u^2$ for $p\approx1$ ($t_2=t_1+\tau$, darker points) and $p=0.60(2)$ ($t_2=t_1+95(5)\,$ms, lighter points) for different integration times $\tau$. For $p\approx 1$, the unexplained variance corresponds to twice the reduced Allan variance (solid orange line). The bright orange line denotes the full Allan variance without subtracting the mean dynamics (see Fig.~\ref{fig:1}e). The gray tick indicates the integration time $\tau=4\,$ms that minimizes $\sigma_u^2$ and was used to produce the curves shown in \textbf{a} and \textbf{b}.  
\textbf{d} Scaling of unexplained variance for fixed lost fraction $p=0.60(2)$ with initial atom number $N_1$. The constant measurement noise has been subtracted from the data for each $n$, to compare different $n$ (color coded as in Fig.~\ref{fig:2}). The gray solid line denotes the linear Poissonian expectation $\tilde{\sigma}_u^2(N_1)=p(1-p)\,N_1$. The errorbars denote one standard deviation.}
\end{figure*} 
\section{Non-linearity of evaporation}
The non-linear dynamics of a system are captured already by the evolution of statistical averages. In the following, we outline how two-time correlations extracted from a continuous measurement of the atom number of the same cloud enable us to uncover the non-linear character of evaporative cooling. To this end, we compare the remaining fraction of atoms, $p=\avg{N_2}/\avg{N_1}$, to the slope extracted from scatter plots such as those shown in Fig.~\ref{fig:2}, calculated as the least squares estimate $a=\rho_{12}\,\sigma_2/\sigma_1$. For a linear process, defined by a constant evaporation rate per atom and hence $\dot{N}\propto -N$, the two quantities should coincide and hence $a=p$, whereas non-linear effects lead to a deviation from this expectation; see Fig.~\ref{fig:3}a and Appendix C.
As an example, the case $\dot{N}\propto -N^{\alpha}$ with $\alpha>1$ leads to $a<p$, which we term ``sub-linear'' to indicate a scatter slope smaller than in the linear case.\\
For our evaporation sequence, we find systematic deviations from a simple linear relationship; see Fig.~\ref{fig:3}b. Recording the difference $a_s = a-p$ versus the remaining fraction of atoms $p$, we observe a pronounced super-linear behavior with $a>p$, when $p$ is calculated relative to a small initial time $t_1$.
We interpret this super-linear behavior as being mediated by temperature variations in our sample distribution:  Atomic gases are prepared with a low initial atom number because of higher atom losses from an initially high gas temperature.  The high temperature hastens evaporation, leading to a higher per-atom loss rate and thus a relatively lower final atom number, leading altogether to $a > p$. This picture is backed by simulations of the evaporation process; see Fig.~\ref{fig:S3} and Appendix B.\\
The rapid evaporation of hotter gases also quickly reduces their temperatures, such that over time different realizations of the gas arrive at nearly the same final temperature irrespective of their initial temperatures.
This interesting transient behavior is often implicitly assumed in literature quoting that the temperature locks to a fixed fraction $1/\eta$ of the trap depth~\cite{Luiten1996a,Ketterle1996a,OHara2001a} and ultimately originates from the competition between exponential truncation and temperature-dependent thermalization of atoms in the evaporation process.\\
At later times in the evaporation process, after the effect of initial temperature variation is suppressed, we find slightly sub-linear behavior, where an initial excess of atoms at time $t_1$ is reduced during the evolution. This behavior can be explained by increased three-body losses towards the end of the evaporation ramp, which reduce ensemble variations due to their non-linear character~\cite{Whitlock2010}; see Fig.~\ref{fig:S3}.
Interestingly, our observations imply that evaporation first moderates temperature variation and then, thereafter, atom number variation in different realizations, ultimately resulting in a stabilizing effect for both atom number and temperature.
%%%%%%%%%%%%%%%%%%%%%%%%%%%%%%%%%%%%%%%%%%%%%%%%%%%%%%%%%%%%%%%%%%%%%%%
%                       	  Figure 4                                %
%%%%%%%%%%%%%%%%%%%%%%%%%%%%%%%%%%%%%%%%%%%%%%%%%%%%%%%%%%%%%%%%%%%%%%%
\section{Stochastic character of evaporation}
The mesoscopic nature of our samples together with the continuous, high-precision measurement enables us to characterize also the stochastic fluctuations inherent in evaporative cooling accrued on a single trajectory. These cannot be obtained from the average evolution of an evaporating gas.
Crucially, analysis of two-time correlations allows for discriminating the effect of variation in initial conditions from the inherent stochastic fluctuations, which reach up to an rms on the order of $\sqrt{N}$ in mesoscopic samples with small $N$.\\
To access stochastic fluctuations directly, we analyze the unexplained variance $\sigma_u^2=\sigma_2^2\left(1-\rho_{12}^2\right)$ as the amount of variance in atom number measured at time $t_2$ that is not predictable through correlation with the initial atom number at time $t_1$; that is, the unexplained variance captures the variance beyond that explained by the linear slope in an $N_1$--$N_2$ scatter plot; see also Figs.~\ref{fig:2} and~\ref{fig:3}.
The unexplained variance also generalizes the reduced Allan variance (see Fig.~\ref{fig:1}e) to arbitrary time separations, and is dominated by the measurement noise and additionally the stochastic noise associated with atom loss; see Appendix C.\\
We observe distinctly different behavior at early times and at late times during the evaporation process. Referenced to an early initial time $t_1 = 10$ ms, the the unexplained variance grows rapidly; see Fig.~\ref{fig:4}a. Compared to a simple parameter-free theoretical model (see Appendix C) we find that the fluctuations are up to a factor of three above the fluctuations expected for a purely uncorrelated atom loss, which is described by a Poissonian stochastic process. Consistent with our earlier interpretation, we ascribe these large fluctuations to additional variance in the initial temperature that is uncorrelated with the initial atom number.\\
Referenced to a later initial time $t_1 = 180$ ms, when unexplained temperature variations have been suppressed, we observe a slower growth of $\sigma_u^2$. Here, we find good agreement with a linear, Poissonian stochastic loss process, indicating that linear single-atom loss dominates the fluctuations of the cooling process even in presence of a non-linear three-body loss; see Fig.~\ref{fig:4}b.
For large remaining fraction of atoms $p\approx1$ ($t_2$ close to $t_1$), the unexplained variance is consistent with the sum of measurement noise in $N_1$ and $N_2$ and, hence, coincides with twice the reduced Allan variance; see Fig.~\ref{fig:4}c. We note that, in the fluctuation analysis, the mean dynamics is subtracted, such that the optimal integration time can be increased to $\tau=4\,$ms without compromising time resolution of our dynamical measurement; see Fig.~\ref{fig:1}e.\\
In order to further corroborate the stochastic nature of the unexplained variance above measurement noise, we investigated the unexplained variance at a fixed fraction of lost atoms $p=0.60(2)$ for varying initial atom numbers $N_1$; see Fig.~\ref{fig:4}d. This strategy is motivated by quantum optics experiments, where the variance originating from photon (quantum) shot noise ($\propto n$) can be discriminated against technical noise ($\propto n^2$) through their scaling with photon number.
In our experiment, we again find a strong excess in $\sigma_u^2$ for large initial atom numbers $N_1$. However, for smaller initial atom numbers, the unexplained variance approaches and traces the Poissonian expectation $p(1-p)\,N_1$ for a range of $N_1$ at all $n$. This contrasts with the quadratic dependence of unexplained variance on initial atom number expected for technical noise.
 
%%%%%%%%%%%%%%%%%%%%%%%%%%%%%%%%%%%%%%%%%%%%%%%%%%%%%%%%%%%%%%%%%%%%%%%%
%%                             Conclusion                              %
%%%%%%%%%%%%%%%%%%%%%%%%%%%%%%%%%%%%%%%%%%%%%%%%%%%%%%%%%%%%%%%%%%%%%%%%
\section{Conclusion}
The minimally invasive measurement of atom number dynamics in our cavity-coupled atomic gas opens numerous near- and longer term perspectives.
First, our work provides an ideal starting point for further studies of evaporation dynamics in low-dimensional mesoscopic quantum gases. Specifically, our results call for a detailed study of the impact of $N$-body losses on evaporation, which are expected to lead to enhanced atom-number fluctuations but also have a moderating effect on ensemble variations due to the non-linear density dependence of the evaporation rate.\\
Our results also indicate that feedback on the atom number~\cite{Gajdacz2013,Gajdacz2016} based on the non-destructive real-time record of an evaporating gas can prepare samples at a fixed temperature with controllable atom numbers fluctuating less than the amount set by Poissonian statistics. Samples stabilized this way provide an ideal starting condition to study atom-number-dependent collective phenomena in optical cavities such as dynamical instabilities~\cite{Kohler2018}.\\
Furthermore, the densities reached in our two-dimensional system are close to the regime where corrections due to Bose statistics and interactions in the gas become relevant. We expect that we can reach this regime with only slightly higher initial atom numbers, which were limited technically in the present study. This prospect motivates further studies focusing on how evaporation dynamics, thermodynamics and stochastic fluctuations are modified by quantum statistics and atomic interactions. Approaching the superfluid transition, our technique complements recent studies in three-dimensional gases~\cite{Kristensen2019,Christensen2021}. In particular, it enables highly sensitive studies of dynamical fluctuation growth and stochastic behavior in a two-dimensional mesoscopic setting and at strong interactions, providing a new window into the intriguing physics of interacting two-dimensional Bose gases~\cite{Hadzibabic2011}.\\  
Combining our technique with locally controlled coupling to the cavity~\cite{Periwal2021} opens the path towards future non-invasive two-terminal transport measurements of strongly correlated quantum gases in optical cavities~\cite{Uchino2018}, for dynamical probing of fluctuation dissipation relations~\cite{Schuckert2020}, or for realizing novel non-destructive local scanning probes of cold gases~\cite{Yang2018}. Such non-destructive probes are also directly relevant to uncovering dynamical fluctuations characteristic of non-equilibrium universality in transport phenomena~\cite{Ferrier2016}, or at fluctuating interfaces~\cite{Halpin-Healy2015}, which can also be studied in neutral-atom systems~\cite{Wei2021}.\\
Finally, leveraging the recent advances in optical tweezer technology, our detection scheme can be extended to the single-atom level through controlled coupling to optical cavities, realizing a versatile quantum sensor. As a direct application, continuous non-destructive probing of single atoms has recently been identified as a key step to enable energy measurements in many-body systems~\cite{Yang2020}, thereby bringing experimental tests of mesoscopic quantum thermodynamics~\cite{Campisi2011} within reach.\\   
%Bibliography

\begin{acknowledgements}
We thank M. Knap, A. Schuckert, J. Moore, A. Avdoshkin and T. Xu for discussions and A. Rubio-Abadal for comments on the manuscript.
We acknowledge support from NSF QLCI program through grant number OMA-2016245 and also NSF grant PHY-1707756, from the AFOSR (award number FA9550-19-1-0328), from ARO
through the MURI program (grant number W911NF-20-1-0136), and from the California Institute for Quantum Entanglement supported by the Multicampus Research Programs and Initiative of the UC Office of the President (Grant No. MRP-19-601445). The contributions of J.A.I. are funded by the Heising-Simons Foundation through grant 2020-2479 and by the NSF under grant number 1707756.
J.Z. thanks the Humboldt Foundation for the support through a Feodor Lynen fellowship.\\ 
\end{acknowledgements}

\bibliography{RealTimeTrackingEvap}

%apsrev4-2.bst 2019-01-14 (MD) hand-edited version of apsrev4-1.bst
%Control: key (0)
%Control: author (8) initials jnrlst
%Control: editor formatted (1) identically to author
%Control: production of article title (0) allowed
%Control: page (0) single
%Control: year (1) truncated
%Control: production of eprint (0) enabled
\begin{thebibliography}{49}%
\makeatletter
\providecommand \@ifxundefined [1]{%
 \@ifx{#1\undefined}
}%
\providecommand \@ifnum [1]{%
 \ifnum #1\expandafter \@firstoftwo
 \else \expandafter \@secondoftwo
 \fi
}%
\providecommand \@ifx [1]{%
 \ifx #1\expandafter \@firstoftwo
 \else \expandafter \@secondoftwo
 \fi
}%
\providecommand \natexlab [1]{#1}%
\providecommand \enquote  [1]{``#1''}%
\providecommand \bibnamefont  [1]{#1}%
\providecommand \bibfnamefont [1]{#1}%
\providecommand \citenamefont [1]{#1}%
\providecommand \href@noop [0]{\@secondoftwo}%
\providecommand \href [0]{\begingroup \@sanitize@url \@href}%
\providecommand \@href[1]{\@@startlink{#1}\@@href}%
\providecommand \@@href[1]{\endgroup#1\@@endlink}%
\providecommand \@sanitize@url [0]{\catcode `\\12\catcode `\$12\catcode
  `\&12\catcode `\#12\catcode `\^12\catcode `\_12\catcode `\%12\relax}%
\providecommand \@@startlink[1]{}%
\providecommand \@@endlink[0]{}%
\providecommand \url  [0]{\begingroup\@sanitize@url \@url }%
\providecommand \@url [1]{\endgroup\@href {#1}{\urlprefix }}%
\providecommand \urlprefix  [0]{URL }%
\providecommand \Eprint [0]{\href }%
\providecommand \doibase [0]{https://doi.org/}%
\providecommand \selectlanguage [0]{\@gobble}%
\providecommand \bibinfo  [0]{\@secondoftwo}%
\providecommand \bibfield  [0]{\@secondoftwo}%
\providecommand \translation [1]{[#1]}%
\providecommand \BibitemOpen [0]{}%
\providecommand \bibitemStop [0]{}%
\providecommand \bibitemNoStop [0]{.\EOS\space}%
\providecommand \EOS [0]{\spacefactor3000\relax}%
\providecommand \BibitemShut  [1]{\csname bibitem#1\endcsname}%
\let\auto@bib@innerbib\@empty
%</preamble>
\bibitem [{\citenamefont {Campisi}\ \emph {et~al.}(2011)\citenamefont
  {Campisi}, \citenamefont {H{\"a}nggi},\ and\ \citenamefont
  {Talkner}}]{Campisi2011}%
  \BibitemOpen
  \bibfield  {author} {\bibinfo {author} {\bibfnamefont {M.}~\bibnamefont
  {Campisi}}, \bibinfo {author} {\bibfnamefont {P.}~\bibnamefont
  {H{\"a}nggi}},\ and\ \bibinfo {author} {\bibfnamefont {P.}~\bibnamefont
  {Talkner}},\ }\bibfield  {title} {\bibinfo {title} {\emph{Colloquium:
  {Quantum fluctuation relations: Foundations and applications}}},\ }\href
  {https://doi.org/10.1103/RevModPhys.83.771} {\bibfield  {journal} {\bibinfo
  {journal} {Rev. Mod. Phys.}\ }\textbf {\bibinfo {volume} {83}},\ \bibinfo
  {pages} {771} (\bibinfo {year} {2011})}\BibitemShut {NoStop}%
\bibitem [{\citenamefont {Campisi}\ \emph {et~al.}(2010)\citenamefont
  {Campisi}, \citenamefont {Talkner},\ and\ \citenamefont
  {H{\"a}nggi}}]{Campisi2010}%
  \BibitemOpen
  \bibfield  {author} {\bibinfo {author} {\bibfnamefont {M.}~\bibnamefont
  {Campisi}}, \bibinfo {author} {\bibfnamefont {P.}~\bibnamefont {Talkner}},\
  and\ \bibinfo {author} {\bibfnamefont {P.}~\bibnamefont {H{\"a}nggi}},\
  }\bibfield  {title} {\bibinfo {title} {\emph{Fluctuation {{Theorems}} for
  {{Continuously Monitored Quantum Fluxes}}}},\ }\href
  {https://doi.org/10.1103/PhysRevLett.105.140601} {\bibfield  {journal}
  {\bibinfo  {journal} {Phys. Rev. Lett.}\ }\textbf {\bibinfo {volume} {105}},\
  \bibinfo {pages} {140601} (\bibinfo {year} {2010})}\BibitemShut {NoStop}%
\bibitem [{\citenamefont {Blanter}\ and\ \citenamefont
  {B{\"u}ttiker}(2000)}]{Blanter2000}%
  \BibitemOpen
  \bibfield  {author} {\bibinfo {author} {\bibfnamefont {Y.~M.}\ \bibnamefont
  {Blanter}}\ and\ \bibinfo {author} {\bibfnamefont {M.}~\bibnamefont
  {B{\"u}ttiker}},\ }\bibfield  {title} {\bibinfo {title} {\emph{Shot noise in
  mesoscopic conductors}},\ }\href
  {https://doi.org/10.1016/S0370-1573(99)00123-4} {\bibfield  {journal}
  {\bibinfo  {journal} {Physics Reports}\ }\textbf {\bibinfo {volume} {336}},\
  \bibinfo {pages} {1} (\bibinfo {year} {2000})}\BibitemShut {NoStop}%
\bibitem [{\citenamefont {Schottky}(1918)}]{Schottky1918}%
  \BibitemOpen
  \bibfield  {author} {\bibinfo {author} {\bibfnamefont {W.}~\bibnamefont
  {Schottky}},\ }\bibfield  {title} {\bibinfo {title} {\emph{{\"U}ber Spontane
  {{Stromschwankungen}} in Verschiedenen {{Elektrizit\"atsleitern}}}},\ }\href
  {https://doi.org/10.1002/andp.19183622304} {\bibfield  {journal} {\bibinfo
  {journal} {Ann. Phys.}\ }\textbf {\bibinfo {volume} {362}},\ \bibinfo {pages}
  {541} (\bibinfo {year} {1918})}\BibitemShut {NoStop}%
\bibitem [{\citenamefont {Gustavsson}\ \emph {et~al.}(2006)\citenamefont
  {Gustavsson}, \citenamefont {Leturcq}, \citenamefont {Simovi{\v c}},
  \citenamefont {Schleser}, \citenamefont {Ihn}, \citenamefont {Studerus},
  \citenamefont {Ensslin}, \citenamefont {Driscoll},\ and\ \citenamefont
  {Gossard}}]{Gustavsson2006}%
  \BibitemOpen
  \bibfield  {author} {\bibinfo {author} {\bibfnamefont {S.}~\bibnamefont
  {Gustavsson}}, \bibinfo {author} {\bibfnamefont {R.}~\bibnamefont {Leturcq}},
  \bibinfo {author} {\bibfnamefont {B.}~\bibnamefont {Simovi{\v c}}}, \bibinfo
  {author} {\bibfnamefont {R.}~\bibnamefont {Schleser}}, \bibinfo {author}
  {\bibfnamefont {T.}~\bibnamefont {Ihn}}, \bibinfo {author} {\bibfnamefont
  {P.}~\bibnamefont {Studerus}}, \bibinfo {author} {\bibfnamefont
  {K.}~\bibnamefont {Ensslin}}, \bibinfo {author} {\bibfnamefont {D.~C.}\
  \bibnamefont {Driscoll}},\ and\ \bibinfo {author} {\bibfnamefont {A.~C.}\
  \bibnamefont {Gossard}},\ }\bibfield  {title} {\bibinfo {title}
  {\emph{Counting {{Statistics}} of {{Single Electron Transport}} in a
  {{Quantum Dot}}}},\ }\href {https://doi.org/10.1103/PhysRevLett.96.076605}
  {\bibfield  {journal} {\bibinfo  {journal} {Phys. Rev. Lett.}\ }\textbf
  {\bibinfo {volume} {96}},\ \bibinfo {pages} {076605} (\bibinfo {year}
  {2006})}\BibitemShut {NoStop}%
\bibitem [{\citenamefont {Brantut}\ \emph {et~al.}(2012)\citenamefont
  {Brantut}, \citenamefont {Meineke}, \citenamefont {Stadler}, \citenamefont
  {Krinner},\ and\ \citenamefont {Esslinger}}]{Brantut2012}%
  \BibitemOpen
  \bibfield  {author} {\bibinfo {author} {\bibfnamefont {J.-P.}\ \bibnamefont
  {Brantut}}, \bibinfo {author} {\bibfnamefont {J.}~\bibnamefont {Meineke}},
  \bibinfo {author} {\bibfnamefont {D.}~\bibnamefont {Stadler}}, \bibinfo
  {author} {\bibfnamefont {S.}~\bibnamefont {Krinner}},\ and\ \bibinfo {author}
  {\bibfnamefont {T.}~\bibnamefont {Esslinger}},\ }\bibfield  {title} {\bibinfo
  {title} {\emph{Conduction of {{Ultracold Fermions Through}} a {{Mesoscopic
  Channel}}}},\ }\href {https://doi.org/10.1126/science.1223175} {\bibfield
  {journal} {\bibinfo  {journal} {Science}\ }\textbf {\bibinfo {volume}
  {337}},\ \bibinfo {pages} {1069} (\bibinfo {year} {2012})}\BibitemShut
  {NoStop}%
\bibitem [{\citenamefont {Brantut}\ \emph {et~al.}(2013)\citenamefont
  {Brantut}, \citenamefont {Grenier}, \citenamefont {Meineke}, \citenamefont
  {Stadler}, \citenamefont {Krinner}, \citenamefont {Kollath}, \citenamefont
  {Esslinger},\ and\ \citenamefont {Georges}}]{Brantut2013}%
  \BibitemOpen
  \bibfield  {author} {\bibinfo {author} {\bibfnamefont {J.-P.}\ \bibnamefont
  {Brantut}}, \bibinfo {author} {\bibfnamefont {C.}~\bibnamefont {Grenier}},
  \bibinfo {author} {\bibfnamefont {J.}~\bibnamefont {Meineke}}, \bibinfo
  {author} {\bibfnamefont {D.}~\bibnamefont {Stadler}}, \bibinfo {author}
  {\bibfnamefont {S.}~\bibnamefont {Krinner}}, \bibinfo {author} {\bibfnamefont
  {C.}~\bibnamefont {Kollath}}, \bibinfo {author} {\bibfnamefont
  {T.}~\bibnamefont {Esslinger}},\ and\ \bibinfo {author} {\bibfnamefont
  {A.}~\bibnamefont {Georges}},\ }\bibfield  {title} {\bibinfo {title} {\emph{A
  {{Thermoelectric Heat Engine}} with {{Ultracold Atoms}}}},\ }\href
  {https://doi.org/10.1126/science.1242308} {\bibfield  {journal} {\bibinfo
  {journal} {Science}\ }\textbf {\bibinfo {volume} {342}},\ \bibinfo {pages}
  {713} (\bibinfo {year} {2013})}\BibitemShut {NoStop}%
\bibitem [{\citenamefont {Chien}\ \emph {et~al.}(2015)\citenamefont {Chien},
  \citenamefont {Peotta},\ and\ \citenamefont {Di~Ventra}}]{Chien2015}%
  \BibitemOpen
  \bibfield  {author} {\bibinfo {author} {\bibfnamefont {C.-C.}\ \bibnamefont
  {Chien}}, \bibinfo {author} {\bibfnamefont {S.}~\bibnamefont {Peotta}},\ and\
  \bibinfo {author} {\bibfnamefont {M.}~\bibnamefont {Di~Ventra}},\ }\bibfield
  {title} {\bibinfo {title} {\emph{Quantum transport in ultracold atoms}},\
  }\href {https://doi.org/10.1038/nphys3531} {\bibfield  {journal} {\bibinfo
  {journal} {Nat. Phys.}\ }\textbf {\bibinfo {volume} {11}},\ \bibinfo {pages}
  {998} (\bibinfo {year} {2015})}\BibitemShut {NoStop}%
\bibitem [{\citenamefont {{Tanji-Suzuki}}\ \emph {et~al.}(2011)\citenamefont
  {{Tanji-Suzuki}}, \citenamefont {Leroux}, \citenamefont {{Schleier-Smith}},
  \citenamefont {Cetina}, \citenamefont {Grier}, \citenamefont {Simon},\ and\
  \citenamefont {Vuleti{\'c}}}]{Tanji-Suzuki2011}%
  \BibitemOpen
  \bibfield  {author} {\bibinfo {author} {\bibfnamefont {H.}~\bibnamefont
  {{Tanji-Suzuki}}}, \bibinfo {author} {\bibfnamefont {I.~D.}\ \bibnamefont
  {Leroux}}, \bibinfo {author} {\bibfnamefont {M.~H.}\ \bibnamefont
  {{Schleier-Smith}}}, \bibinfo {author} {\bibfnamefont {M.}~\bibnamefont
  {Cetina}}, \bibinfo {author} {\bibfnamefont {A.~T.}\ \bibnamefont {Grier}},
  \bibinfo {author} {\bibfnamefont {J.}~\bibnamefont {Simon}},\ and\ \bibinfo
  {author} {\bibfnamefont {V.}~\bibnamefont {Vuleti{\'c}}},\ }\bibfield
  {title} {\bibinfo {title} {\emph{Chapter 4 - {{Interaction}} between {{Atomic
  Ensembles}} and {{Optical Resonators}}: {{Classical Description}}}},\ }\href
  {https://doi.org/10.1016/B978-0-12-385508-4.00004-8} {\bibfield  {journal}
  {\bibinfo  {journal} {Advances {{In Atomic}}, {{Molecular}}, and {{Optical
  Physics}}}\ }\textbf {\bibinfo {volume} {60}},\ \bibinfo {pages} {201}
  (\bibinfo {year} {2011})}\BibitemShut {NoStop}%
\bibitem [{\citenamefont {Hood}\ \emph {et~al.}(2000)\citenamefont {Hood},
  \citenamefont {Lynn}, \citenamefont {Doherty}, \citenamefont {Parkins},\ and\
  \citenamefont {Kimble}}]{Hood2000}%
  \BibitemOpen
  \bibfield  {author} {\bibinfo {author} {\bibfnamefont {C.~J.}\ \bibnamefont
  {Hood}}, \bibinfo {author} {\bibfnamefont {T.~W.}\ \bibnamefont {Lynn}},
  \bibinfo {author} {\bibfnamefont {A.~C.}\ \bibnamefont {Doherty}}, \bibinfo
  {author} {\bibfnamefont {A.~S.}\ \bibnamefont {Parkins}},\ and\ \bibinfo
  {author} {\bibfnamefont {H.~J.}\ \bibnamefont {Kimble}},\ }\bibfield  {title}
  {\bibinfo {title} {\emph{The {{Atom}}-{{Cavity Microscope}}: {{Single Atoms
  Bound}} in {{Orbit}} by {{Single Photons}}}},\ }\href
  {https://doi.org/10.1126/science.287.5457.1447} {\bibfield  {journal}
  {\bibinfo  {journal} {Science}\ }\textbf {\bibinfo {volume} {287}},\ \bibinfo
  {pages} {1447} (\bibinfo {year} {2000})}\BibitemShut {NoStop}%
\bibitem [{\citenamefont {Pinkse}\ \emph {et~al.}(2000)\citenamefont {Pinkse},
  \citenamefont {Fischer}, \citenamefont {Maunz},\ and\ \citenamefont
  {Rempe}}]{Pinkse2000}%
  \BibitemOpen
  \bibfield  {author} {\bibinfo {author} {\bibfnamefont {P.~W.~H.}\
  \bibnamefont {Pinkse}}, \bibinfo {author} {\bibfnamefont {T.}~\bibnamefont
  {Fischer}}, \bibinfo {author} {\bibfnamefont {P.}~\bibnamefont {Maunz}},\
  and\ \bibinfo {author} {\bibfnamefont {G.}~\bibnamefont {Rempe}},\ }\bibfield
   {title} {\bibinfo {title} {\emph{Trapping an atom with single photons}},\
  }\href {https://doi.org/10.1038/35006006} {\bibfield  {journal} {\bibinfo
  {journal} {Nature}\ }\textbf {\bibinfo {volume} {404}},\ \bibinfo {pages}
  {365} (\bibinfo {year} {2000})}\BibitemShut {NoStop}%
\bibitem [{\citenamefont {Brahms}\ \emph {et~al.}(2011)\citenamefont {Brahms},
  \citenamefont {Purdy}, \citenamefont {Brooks}, \citenamefont {Botter},\ and\
  \citenamefont {{Stamper-Kurn}}}]{Brahms2011}%
  \BibitemOpen
  \bibfield  {author} {\bibinfo {author} {\bibfnamefont {N.}~\bibnamefont
  {Brahms}}, \bibinfo {author} {\bibfnamefont {T.~P.}\ \bibnamefont {Purdy}},
  \bibinfo {author} {\bibfnamefont {D.~W.~C.}\ \bibnamefont {Brooks}}, \bibinfo
  {author} {\bibfnamefont {T.}~\bibnamefont {Botter}},\ and\ \bibinfo {author}
  {\bibfnamefont {D.~M.}\ \bibnamefont {{Stamper-Kurn}}},\ }\bibfield  {title}
  {\bibinfo {title} {\emph{Cavity-aided magnetic resonance microscopy of atomic
  transport in optical lattices}},\ }\href {https://doi.org/10.1038/nphys1967}
  {\bibfield  {journal} {\bibinfo  {journal} {Nat. Phys.}\ }\textbf {\bibinfo
  {volume} {7}},\ \bibinfo {pages} {604} (\bibinfo {year} {2011})}\BibitemShut
  {NoStop}%
\bibitem [{\citenamefont {Zhang}\ \emph {et~al.}(2012)\citenamefont {Zhang},
  \citenamefont {McConnell}, \citenamefont {{\'C}uk}, \citenamefont {Lin},
  \citenamefont {{Schleier-Smith}}, \citenamefont {Leroux},\ and\ \citenamefont
  {Vuleti{\'c}}}]{Zhang2012}%
  \BibitemOpen
  \bibfield  {author} {\bibinfo {author} {\bibfnamefont {H.}~\bibnamefont
  {Zhang}}, \bibinfo {author} {\bibfnamefont {R.}~\bibnamefont {McConnell}},
  \bibinfo {author} {\bibfnamefont {S.}~\bibnamefont {{\'C}uk}}, \bibinfo
  {author} {\bibfnamefont {Q.}~\bibnamefont {Lin}}, \bibinfo {author}
  {\bibfnamefont {M.~H.}\ \bibnamefont {{Schleier-Smith}}}, \bibinfo {author}
  {\bibfnamefont {I.~D.}\ \bibnamefont {Leroux}},\ and\ \bibinfo {author}
  {\bibfnamefont {V.}~\bibnamefont {Vuleti{\'c}}},\ }\bibfield  {title}
  {\bibinfo {title} {\emph{Collective {{State Measurement}} of {{Mesoscopic
  Ensembles}} with {{Single}}-{{Atom Resolution}}}},\ }\href
  {https://doi.org/10.1103/PhysRevLett.109.133603} {\bibfield  {journal}
  {\bibinfo  {journal} {Phys. Rev. Lett.}\ }\textbf {\bibinfo {volume} {109}},\
  \bibinfo {pages} {133603} (\bibinfo {year} {2012})}\BibitemShut {NoStop}%
\bibitem [{\citenamefont {Chen}\ \emph {et~al.}(2014)\citenamefont {Chen},
  \citenamefont {Bohnet}, \citenamefont {Weiner}, \citenamefont {Cox},\ and\
  \citenamefont {Thompson}}]{Chen2014}%
  \BibitemOpen
  \bibfield  {author} {\bibinfo {author} {\bibfnamefont {Z.}~\bibnamefont
  {Chen}}, \bibinfo {author} {\bibfnamefont {J.~G.}\ \bibnamefont {Bohnet}},
  \bibinfo {author} {\bibfnamefont {J.~M.}\ \bibnamefont {Weiner}}, \bibinfo
  {author} {\bibfnamefont {K.~C.}\ \bibnamefont {Cox}},\ and\ \bibinfo {author}
  {\bibfnamefont {J.~K.}\ \bibnamefont {Thompson}},\ }\bibfield  {title}
  {\bibinfo {title} {\emph{Cavity-aided nondemolition measurements for atom
  counting and spin squeezing}},\ }\href
  {https://doi.org/10.1103/PhysRevA.89.043837} {\bibfield  {journal} {\bibinfo
  {journal} {Phys. Rev. A}\ }\textbf {\bibinfo {volume} {89}},\ \bibinfo
  {pages} {043837} (\bibinfo {year} {2014})}\BibitemShut {NoStop}%
\bibitem [{\citenamefont {Norcia}\ and\ \citenamefont
  {Thompson}(2016)}]{Norcia2016}%
  \BibitemOpen
  \bibfield  {author} {\bibinfo {author} {\bibfnamefont {M.~A.}\ \bibnamefont
  {Norcia}}\ and\ \bibinfo {author} {\bibfnamefont {J.~K.}\ \bibnamefont
  {Thompson}},\ }\bibfield  {title} {\bibinfo {title} {\emph{Strong coupling on
  a forbidden transition in strontium and nondestructive atom counting}},\
  }\href {https://doi.org/10.1103/PhysRevA.93.023804} {\bibfield  {journal}
  {\bibinfo  {journal} {Phys. Rev. A}\ }\textbf {\bibinfo {volume} {93}},\
  \bibinfo {pages} {023804} (\bibinfo {year} {2016})}\BibitemShut {NoStop}%
\bibitem [{\citenamefont {Kohler}\ \emph {et~al.}(2017)\citenamefont {Kohler},
  \citenamefont {Spethmann}, \citenamefont {Schreppler},\ and\ \citenamefont
  {{Stamper-Kurn}}}]{Kohler2017}%
  \BibitemOpen
  \bibfield  {author} {\bibinfo {author} {\bibfnamefont {J.}~\bibnamefont
  {Kohler}}, \bibinfo {author} {\bibfnamefont {N.}~\bibnamefont {Spethmann}},
  \bibinfo {author} {\bibfnamefont {S.}~\bibnamefont {Schreppler}},\ and\
  \bibinfo {author} {\bibfnamefont {D.~M.}\ \bibnamefont {{Stamper-Kurn}}},\
  }\bibfield  {title} {\bibinfo {title} {\emph{Cavity-{{Assisted Measurement}}
  and {{Coherent Control}} of {{Collective Atomic Spin Oscillators}}}},\ }\href
  {https://doi.org/10.1103/PhysRevLett.118.063604} {\bibfield  {journal}
  {\bibinfo  {journal} {Phys. Rev. Lett.}\ }\textbf {\bibinfo {volume} {118}},\
  \bibinfo {pages} {063604} (\bibinfo {year} {2017})}\BibitemShut {NoStop}%
\bibitem [{\citenamefont {Roux}\ \emph {et~al.}(2021)\citenamefont {Roux},
  \citenamefont {Helson}, \citenamefont {Konishi},\ and\ \citenamefont
  {Brantut}}]{Roux2021}%
  \BibitemOpen
  \bibfield  {author} {\bibinfo {author} {\bibfnamefont {K.}~\bibnamefont
  {Roux}}, \bibinfo {author} {\bibfnamefont {V.}~\bibnamefont {Helson}},
  \bibinfo {author} {\bibfnamefont {H.}~\bibnamefont {Konishi}},\ and\ \bibinfo
  {author} {\bibfnamefont {J.-P.}\ \bibnamefont {Brantut}},\ }\bibfield
  {title} {\bibinfo {title} {\emph{Cavity-assisted preparation and detection of
  a unitary {{Fermi}} gas}},\ }\href {https://doi.org/10.1088/1367-2630/abeb91}
  {\bibfield  {journal} {\bibinfo  {journal} {New J. Phys.}\ }\textbf {\bibinfo
  {volume} {23}},\ \bibinfo {pages} {043029} (\bibinfo {year}
  {2021})}\BibitemShut {NoStop}%
\bibitem [{\citenamefont {L{\'e}onard}\ \emph
  {et~al.}(2017{\natexlab{a}})\citenamefont {L{\'e}onard}, \citenamefont
  {Morales}, \citenamefont {Zupancic}, \citenamefont {Esslinger},\ and\
  \citenamefont {Donner}}]{Leonard2017c}%
  \BibitemOpen
  \bibfield  {author} {\bibinfo {author} {\bibfnamefont {J.}~\bibnamefont
  {L{\'e}onard}}, \bibinfo {author} {\bibfnamefont {A.}~\bibnamefont
  {Morales}}, \bibinfo {author} {\bibfnamefont {P.}~\bibnamefont {Zupancic}},
  \bibinfo {author} {\bibfnamefont {T.}~\bibnamefont {Esslinger}},\ and\
  \bibinfo {author} {\bibfnamefont {T.}~\bibnamefont {Donner}},\ }\bibfield
  {title} {\bibinfo {title} {\emph{Supersolid formation in a quantum gas
  breaking a continuous translational symmetry}},\ }\href
  {https://doi.org/10.1038/nature21067} {\bibfield  {journal} {\bibinfo
  {journal} {Nature}\ }\textbf {\bibinfo {volume} {543}},\ \bibinfo {pages}
  {87} (\bibinfo {year} {2017}{\natexlab{a}})}\BibitemShut {NoStop}%
\bibitem [{\citenamefont {L{\'e}onard}\ \emph
  {et~al.}(2017{\natexlab{b}})\citenamefont {L{\'e}onard}, \citenamefont
  {Morales}, \citenamefont {Zupancic}, \citenamefont {Donner},\ and\
  \citenamefont {Esslinger}}]{Leonard2017b}%
  \BibitemOpen
  \bibfield  {author} {\bibinfo {author} {\bibfnamefont {J.}~\bibnamefont
  {L{\'e}onard}}, \bibinfo {author} {\bibfnamefont {A.}~\bibnamefont
  {Morales}}, \bibinfo {author} {\bibfnamefont {P.}~\bibnamefont {Zupancic}},
  \bibinfo {author} {\bibfnamefont {T.}~\bibnamefont {Donner}},\ and\ \bibinfo
  {author} {\bibfnamefont {T.}~\bibnamefont {Esslinger}},\ }\bibfield  {title}
  {\bibinfo {title} {\emph{Monitoring and manipulating {{Higgs}} and
  {{Goldstone}} modes in a supersolid quantum gas}},\ }\href
  {https://doi.org/10.1126/science.aan2608} {\bibfield  {journal} {\bibinfo
  {journal} {Science}\ }\textbf {\bibinfo {volume} {358}},\ \bibinfo {pages}
  {1415} (\bibinfo {year} {2017}{\natexlab{b}})}\BibitemShut {NoStop}%
\bibitem [{\citenamefont {Kroeze}\ \emph {et~al.}(2019)\citenamefont {Kroeze},
  \citenamefont {Guo},\ and\ \citenamefont {Lev}}]{Kroeze2019}%
  \BibitemOpen
  \bibfield  {author} {\bibinfo {author} {\bibfnamefont {R.~M.}\ \bibnamefont
  {Kroeze}}, \bibinfo {author} {\bibfnamefont {Y.}~\bibnamefont {Guo}},\ and\
  \bibinfo {author} {\bibfnamefont {B.~L.}\ \bibnamefont {Lev}},\ }\bibfield
  {title} {\bibinfo {title} {\emph{Dynamical {{Spin}}-{{Orbit Coupling}} of a
  {{Quantum Gas}}}},\ }\href {https://doi.org/10.1103/PhysRevLett.123.160404}
  {\bibfield  {journal} {\bibinfo  {journal} {Phys. Rev. Lett.}\ }\textbf
  {\bibinfo {volume} {123}},\ \bibinfo {pages} {160404} (\bibinfo {year}
  {2019})}\BibitemShut {NoStop}%
\bibitem [{\citenamefont {Krinner}\ \emph {et~al.}(2017)\citenamefont
  {Krinner}, \citenamefont {Esslinger},\ and\ \citenamefont
  {Brantut}}]{Krinner2017}%
  \BibitemOpen
  \bibfield  {author} {\bibinfo {author} {\bibfnamefont {S.}~\bibnamefont
  {Krinner}}, \bibinfo {author} {\bibfnamefont {T.}~\bibnamefont {Esslinger}},\
  and\ \bibinfo {author} {\bibfnamefont {J.-P.}\ \bibnamefont {Brantut}},\
  }\bibfield  {title} {\bibinfo {title} {\emph{Two-terminal transport
  measurements with cold atoms}},\ }\href
  {https://doi.org/10.1088/1361-648X/aa74a1} {\bibfield  {journal} {\bibinfo
  {journal} {J. Phys.: Condens. Matter}\ }\textbf {\bibinfo {volume} {29}},\
  \bibinfo {pages} {343003} (\bibinfo {year} {2017})}\BibitemShut {NoStop}%
\bibitem [{\citenamefont {Uchino}\ \emph {et~al.}(2018)\citenamefont {Uchino},
  \citenamefont {Ueda},\ and\ \citenamefont {Brantut}}]{Uchino2018}%
  \BibitemOpen
  \bibfield  {author} {\bibinfo {author} {\bibfnamefont {S.}~\bibnamefont
  {Uchino}}, \bibinfo {author} {\bibfnamefont {M.}~\bibnamefont {Ueda}},\ and\
  \bibinfo {author} {\bibfnamefont {J.-P.}\ \bibnamefont {Brantut}},\
  }\bibfield  {title} {\bibinfo {title} {\emph{Universal noise in continuous
  transport measurements of interacting fermions}},\ }\href
  {https://doi.org/10.1103/PhysRevA.98.063619} {\bibfield  {journal} {\bibinfo
  {journal} {Phys. Rev. A}\ }\textbf {\bibinfo {volume} {98}},\ \bibinfo
  {pages} {063619} (\bibinfo {year} {2018})}\BibitemShut {NoStop}%
\bibitem [{\citenamefont {Yang}\ \emph {et~al.}(2018)\citenamefont {Yang},
  \citenamefont {Laflamme}, \citenamefont {Vasilyev}, \citenamefont {Baranov},\
  and\ \citenamefont {Zoller}}]{Yang2018}%
  \BibitemOpen
  \bibfield  {author} {\bibinfo {author} {\bibfnamefont {D.}~\bibnamefont
  {Yang}}, \bibinfo {author} {\bibfnamefont {C.}~\bibnamefont {Laflamme}},
  \bibinfo {author} {\bibfnamefont {D.~V.}\ \bibnamefont {Vasilyev}}, \bibinfo
  {author} {\bibfnamefont {M.~A.}\ \bibnamefont {Baranov}},\ and\ \bibinfo
  {author} {\bibfnamefont {P.}~\bibnamefont {Zoller}},\ }\bibfield  {title}
  {\bibinfo {title} {\emph{Theory of a {{Quantum Scanning Microscope}} for
  {{Cold Atoms}}}},\ }\href {https://doi.org/10.1103/PhysRevLett.120.133601}
  {\bibfield  {journal} {\bibinfo  {journal} {Phys. Rev. Lett.}\ }\textbf
  {\bibinfo {volume} {120}},\ \bibinfo {pages} {133601} (\bibinfo {year}
  {2018})}\BibitemShut {NoStop}%
\bibitem [{\citenamefont {Sawyer}\ \emph {et~al.}(2012)\citenamefont {Sawyer},
  \citenamefont {Deb}, \citenamefont {McKellar},\ and\ \citenamefont
  {Kj{\ae}rgaard}}]{Sawyer2012}%
  \BibitemOpen
  \bibfield  {author} {\bibinfo {author} {\bibfnamefont {B.~J.}\ \bibnamefont
  {Sawyer}}, \bibinfo {author} {\bibfnamefont {A.~B.}\ \bibnamefont {Deb}},
  \bibinfo {author} {\bibfnamefont {T.}~\bibnamefont {McKellar}},\ and\
  \bibinfo {author} {\bibfnamefont {N.}~\bibnamefont {Kj{\ae}rgaard}},\
  }\bibfield  {title} {\bibinfo {title} {\emph{Reducing number fluctuations of
  ultracold atomic gases via dispersive interrogation}},\ }\href
  {https://doi.org/10.1103/PhysRevA.86.065401} {\bibfield  {journal} {\bibinfo
  {journal} {Phys. Rev. A}\ }\textbf {\bibinfo {volume} {86}},\ \bibinfo
  {pages} {065401} (\bibinfo {year} {2012})}\BibitemShut {NoStop}%
\bibitem [{\citenamefont {Gajdacz}\ \emph {et~al.}(2013)\citenamefont
  {Gajdacz}, \citenamefont {Pedersen}, \citenamefont {M{\o}rch}, \citenamefont
  {Hilliard}, \citenamefont {Arlt},\ and\ \citenamefont
  {Sherson}}]{Gajdacz2013}%
  \BibitemOpen
  \bibfield  {author} {\bibinfo {author} {\bibfnamefont {M.}~\bibnamefont
  {Gajdacz}}, \bibinfo {author} {\bibfnamefont {P.~L.}\ \bibnamefont
  {Pedersen}}, \bibinfo {author} {\bibfnamefont {T.}~\bibnamefont {M{\o}rch}},
  \bibinfo {author} {\bibfnamefont {A.~J.}\ \bibnamefont {Hilliard}}, \bibinfo
  {author} {\bibfnamefont {J.}~\bibnamefont {Arlt}},\ and\ \bibinfo {author}
  {\bibfnamefont {J.~F.}\ \bibnamefont {Sherson}},\ }\bibfield  {title}
  {\bibinfo {title} {\emph{Non-destructive {{Faraday}} imaging of dynamically
  controlled ultracold atoms}},\ }\href {https://doi.org/10.1063/1.4818913}
  {\bibfield  {journal} {\bibinfo  {journal} {Review of Scientific
  Instruments}\ }\textbf {\bibinfo {volume} {84}},\ \bibinfo {pages} {083105}
  (\bibinfo {year} {2013})}\BibitemShut {NoStop}%
\bibitem [{\citenamefont {Kristensen}\ \emph {et~al.}(2019)\citenamefont
  {Kristensen}, \citenamefont {Christensen}, \citenamefont {Gajdacz},
  \citenamefont {Iglicki}, \citenamefont {Paw{\l}owski}, \citenamefont
  {Klempt}, \citenamefont {Sherson}, \citenamefont {Rza{\.z}ewski},
  \citenamefont {Hilliard},\ and\ \citenamefont {Arlt}}]{Kristensen2019}%
  \BibitemOpen
  \bibfield  {author} {\bibinfo {author} {\bibfnamefont {M.~A.}\ \bibnamefont
  {Kristensen}}, \bibinfo {author} {\bibfnamefont {M.~B.}\ \bibnamefont
  {Christensen}}, \bibinfo {author} {\bibfnamefont {M.}~\bibnamefont
  {Gajdacz}}, \bibinfo {author} {\bibfnamefont {M.}~\bibnamefont {Iglicki}},
  \bibinfo {author} {\bibfnamefont {K.}~\bibnamefont {Paw{\l}owski}}, \bibinfo
  {author} {\bibfnamefont {C.}~\bibnamefont {Klempt}}, \bibinfo {author}
  {\bibfnamefont {J.~F.}\ \bibnamefont {Sherson}}, \bibinfo {author}
  {\bibfnamefont {K.}~\bibnamefont {Rza{\.z}ewski}}, \bibinfo {author}
  {\bibfnamefont {A.~J.}\ \bibnamefont {Hilliard}},\ and\ \bibinfo {author}
  {\bibfnamefont {J.~J.}\ \bibnamefont {Arlt}},\ }\bibfield  {title} {\bibinfo
  {title} {\emph{Observation of {{Atom Number Fluctuations}} in a
  {{Bose}}-{{Einstein Condensate}}}},\ }\href
  {https://doi.org/10.1103/PhysRevLett.122.163601} {\bibfield  {journal}
  {\bibinfo  {journal} {Phys. Rev. Lett.}\ }\textbf {\bibinfo {volume} {122}},\
  \bibinfo {pages} {163601} (\bibinfo {year} {2019})}\BibitemShut {NoStop}%
\bibitem [{\citenamefont {Christensen}\ \emph {et~al.}(2021)\citenamefont
  {Christensen}, \citenamefont {Vibel}, \citenamefont {Hilliard}, \citenamefont
  {Kruk}, \citenamefont {Paw{\l}owski}, \citenamefont {Hryniuk}, \citenamefont
  {Rza{\.z}ewski}, \citenamefont {Kristensen},\ and\ \citenamefont
  {Arlt}}]{Christensen2021}%
  \BibitemOpen
  \bibfield  {author} {\bibinfo {author} {\bibfnamefont {M.~B.}\ \bibnamefont
  {Christensen}}, \bibinfo {author} {\bibfnamefont {T.}~\bibnamefont {Vibel}},
  \bibinfo {author} {\bibfnamefont {A.~J.}\ \bibnamefont {Hilliard}}, \bibinfo
  {author} {\bibfnamefont {M.~B.}\ \bibnamefont {Kruk}}, \bibinfo {author}
  {\bibfnamefont {K.}~\bibnamefont {Paw{\l}owski}}, \bibinfo {author}
  {\bibfnamefont {D.}~\bibnamefont {Hryniuk}}, \bibinfo {author} {\bibfnamefont
  {K.}~\bibnamefont {Rza{\.z}ewski}}, \bibinfo {author} {\bibfnamefont {M.~A.}\
  \bibnamefont {Kristensen}},\ and\ \bibinfo {author} {\bibfnamefont {J.~J.}\
  \bibnamefont {Arlt}},\ }\bibfield  {title} {\bibinfo {title}
  {\emph{Observation of {{Microcanonical Atom Number Fluctuations}} in a
  {{Bose}}-{{Einstein Condensate}}}},\ }\href
  {https://doi.org/10.1103/PhysRevLett.126.153601} {\bibfield  {journal}
  {\bibinfo  {journal} {Phys. Rev. Lett.}\ }\textbf {\bibinfo {volume} {126}},\
  \bibinfo {pages} {153601} (\bibinfo {year} {2021})}\BibitemShut {NoStop}%
\bibitem [{\citenamefont {Davis}\ \emph {et~al.}(1995)\citenamefont {Davis},
  \citenamefont {Mewes},\ and\ \citenamefont {Ketterle}}]{Davis1995b}%
  \BibitemOpen
  \bibfield  {author} {\bibinfo {author} {\bibfnamefont {K.~B.}\ \bibnamefont
  {Davis}}, \bibinfo {author} {\bibfnamefont {M.~O.}\ \bibnamefont {Mewes}},\
  and\ \bibinfo {author} {\bibfnamefont {W.}~\bibnamefont {Ketterle}},\
  }\bibfield  {title} {\bibinfo {title} {\emph{An analytical model for
  evaporative cooling of atoms}},\ }\href {https://doi.org/10.1007/BF01135857}
  {\bibfield  {journal} {\bibinfo  {journal} {Appl. Phys. B}\ }\textbf
  {\bibinfo {volume} {60}},\ \bibinfo {pages} {155} (\bibinfo {year}
  {1995})}\BibitemShut {NoStop}%
\bibitem [{\citenamefont {Luiten}\ \emph {et~al.}(1996)\citenamefont {Luiten},
  \citenamefont {Reynolds},\ and\ \citenamefont {Walraven}}]{Luiten1996a}%
  \BibitemOpen
  \bibfield  {author} {\bibinfo {author} {\bibfnamefont {O.~J.}\ \bibnamefont
  {Luiten}}, \bibinfo {author} {\bibfnamefont {M.~W.}\ \bibnamefont
  {Reynolds}},\ and\ \bibinfo {author} {\bibfnamefont {J.~T.~M.}\ \bibnamefont
  {Walraven}},\ }\bibfield  {title} {\bibinfo {title} {\emph{Kinetic theory of
  the evaporative cooling of a trapped gas}},\ }\href
  {https://doi.org/10.1103/PhysRevA.53.381} {\bibfield  {journal} {\bibinfo
  {journal} {Phys. Rev. A}\ }\textbf {\bibinfo {volume} {53}},\ \bibinfo
  {pages} {381} (\bibinfo {year} {1996})}\BibitemShut {NoStop}%
\bibitem [{\citenamefont {Ketterle}\ and\ \citenamefont
  {Druten}(1996)}]{Ketterle1996a}%
  \BibitemOpen
  \bibfield  {author} {\bibinfo {author} {\bibfnamefont {W.}~\bibnamefont
  {Ketterle}}\ and\ \bibinfo {author} {\bibfnamefont {N.~J.~V.}\ \bibnamefont
  {Druten}},\ }\bibfield  {title} {\bibinfo {title} {\emph{Evaporative
  {{Cooling}} of {{Trapped Atoms}}}},\ }\href
  {https://doi.org/10.1016/S1049-250X(08)60101-9} {\bibfield  {journal}
  {\bibinfo  {journal} {Advances {{In Atomic}}, {{Molecular}}, and {{Optical
  Physics}}}\ }\textbf {\bibinfo {volume} {37}},\ \bibinfo {pages} {181}
  (\bibinfo {year} {1996})}\BibitemShut {NoStop}%
\bibitem [{\citenamefont {Wu}\ and\ \citenamefont {Foot}(1996)}]{Wu1996}%
  \BibitemOpen
  \bibfield  {author} {\bibinfo {author} {\bibfnamefont {H.}~\bibnamefont
  {Wu}}\ and\ \bibinfo {author} {\bibfnamefont {C.~J.}\ \bibnamefont {Foot}},\
  }\bibfield  {title} {\bibinfo {title} {\emph{Direct simulation of evaporative
  cooling}},\ }\href {https://doi.org/10.1088/0953-4075/29/8/003} {\bibfield
  {journal} {\bibinfo  {journal} {J. Phys. B: At. Mol. Opt. Phys.}\ }\textbf
  {\bibinfo {volume} {29}},\ \bibinfo {pages} {L321} (\bibinfo {year}
  {1996})}\BibitemShut {NoStop}%
\bibitem [{\citenamefont {O'Hara}\ \emph {et~al.}(2001)\citenamefont {O'Hara},
  \citenamefont {Gehm}, \citenamefont {Granade},\ and\ \citenamefont
  {Thomas}}]{OHara2001a}%
  \BibitemOpen
  \bibfield  {author} {\bibinfo {author} {\bibfnamefont {K.~M.}\ \bibnamefont
  {O'Hara}}, \bibinfo {author} {\bibfnamefont {M.~E.}\ \bibnamefont {Gehm}},
  \bibinfo {author} {\bibfnamefont {S.~R.}\ \bibnamefont {Granade}},\ and\
  \bibinfo {author} {\bibfnamefont {J.~E.}\ \bibnamefont {Thomas}},\ }\bibfield
   {title} {\bibinfo {title} {\emph{Scaling laws for evaporative cooling in
  time-dependent optical traps}},\ }\href
  {https://doi.org/10.1103/PhysRevA.64.051403} {\bibfield  {journal} {\bibinfo
  {journal} {Phys. Rev. A}\ }\textbf {\bibinfo {volume} {64}},\ \bibinfo
  {pages} {051403(R)} (\bibinfo {year} {2001})}\BibitemShut {NoStop}%
\bibitem [{\citenamefont {Whitlock}\ \emph {et~al.}(2010)\citenamefont
  {Whitlock}, \citenamefont {Ockeloen},\ and\ \citenamefont
  {Spreeuw}}]{Whitlock2010}%
  \BibitemOpen
  \bibfield  {author} {\bibinfo {author} {\bibfnamefont {S.}~\bibnamefont
  {Whitlock}}, \bibinfo {author} {\bibfnamefont {C.~F.}\ \bibnamefont
  {Ockeloen}},\ and\ \bibinfo {author} {\bibfnamefont {R.~J.~C.}\ \bibnamefont
  {Spreeuw}},\ }\bibfield  {title} {\bibinfo {title} {\emph{Sub-{{Poissonian
  Atom}}-{{Number Fluctuations}} by {{Three}}-{{Body Loss}} in {{Mesoscopic
  Ensembles}}}},\ }\href {https://doi.org/10.1103/PhysRevLett.104.120402}
  {\bibfield  {journal} {\bibinfo  {journal} {Phys. Rev. Lett.}\ }\textbf
  {\bibinfo {volume} {104}},\ \bibinfo {pages} {120402} (\bibinfo {year}
  {2010})}\BibitemShut {NoStop}%
\bibitem [{\citenamefont {Kohler}\ \emph {et~al.}(2018)\citenamefont {Kohler},
  \citenamefont {Gerber}, \citenamefont {Dowd},\ and\ \citenamefont
  {{Stamper-Kurn}}}]{Kohler2018}%
  \BibitemOpen
  \bibfield  {author} {\bibinfo {author} {\bibfnamefont {J.}~\bibnamefont
  {Kohler}}, \bibinfo {author} {\bibfnamefont {J.~A.}\ \bibnamefont {Gerber}},
  \bibinfo {author} {\bibfnamefont {E.}~\bibnamefont {Dowd}},\ and\ \bibinfo
  {author} {\bibfnamefont {D.~M.}\ \bibnamefont {{Stamper-Kurn}}},\ }\bibfield
  {title} {\bibinfo {title} {\emph{Negative-{{Mass Instability}} of the
  {{Spin}} and {{Motion}} of an {{Atomic Gas Driven}} by {{Optical Cavity
  Backaction}}}},\ }\href {https://doi.org/10.1103/PhysRevLett.120.013601}
  {\bibfield  {journal} {\bibinfo  {journal} {Phys. Rev. Lett.}\ }\textbf
  {\bibinfo {volume} {120}},\ \bibinfo {pages} {013601} (\bibinfo {year}
  {2018})}\BibitemShut {NoStop}%
\bibitem [{\citenamefont {Hung}\ \emph {et~al.}(2008)\citenamefont {Hung},
  \citenamefont {Zhang}, \citenamefont {Gemelke},\ and\ \citenamefont
  {Chin}}]{Hung2008a}%
  \BibitemOpen
  \bibfield  {author} {\bibinfo {author} {\bibfnamefont {C.-L.}\ \bibnamefont
  {Hung}}, \bibinfo {author} {\bibfnamefont {X.}~\bibnamefont {Zhang}},
  \bibinfo {author} {\bibfnamefont {N.}~\bibnamefont {Gemelke}},\ and\ \bibinfo
  {author} {\bibfnamefont {C.}~\bibnamefont {Chin}},\ }\bibfield  {title}
  {\bibinfo {title} {\emph{Accelerating evaporative cooling of atoms into
  {{Bose}}-{{Einstein}} condensation in optical traps}},\ }\href
  {https://doi.org/10.1103/PhysRevA.78.011604} {\bibfield  {journal} {\bibinfo
  {journal} {Phys. Rev. A}\ }\textbf {\bibinfo {volume} {78}},\ \bibinfo
  {pages} {011604(R)} (\bibinfo {year} {2008})}\BibitemShut {NoStop}%
\bibitem [{\citenamefont {Purdy}\ \emph {et~al.}(2010)\citenamefont {Purdy},
  \citenamefont {Brooks}, \citenamefont {Botter}, \citenamefont {Brahms},
  \citenamefont {Ma},\ and\ \citenamefont {{Stamper-Kurn}}}]{Purdy2010a}%
  \BibitemOpen
  \bibfield  {author} {\bibinfo {author} {\bibfnamefont {T.~P.}\ \bibnamefont
  {Purdy}}, \bibinfo {author} {\bibfnamefont {D.~W.~C.}\ \bibnamefont
  {Brooks}}, \bibinfo {author} {\bibfnamefont {T.}~\bibnamefont {Botter}},
  \bibinfo {author} {\bibfnamefont {N.}~\bibnamefont {Brahms}}, \bibinfo
  {author} {\bibfnamefont {Z.-Y.}\ \bibnamefont {Ma}},\ and\ \bibinfo {author}
  {\bibfnamefont {D.~M.}\ \bibnamefont {{Stamper-Kurn}}},\ }\bibfield  {title}
  {\bibinfo {title} {\emph{Tunable {{Cavity Optomechanics}} with {{Ultracold
  Atoms}}}},\ }\href {https://doi.org/10.1103/PhysRevLett.105.133602}
  {\bibfield  {journal} {\bibinfo  {journal} {Phys. Rev. Lett.}\ }\textbf
  {\bibinfo {volume} {105}},\ \bibinfo {pages} {133602} (\bibinfo {year}
  {2010})}\BibitemShut {NoStop}%
\bibitem [{\citenamefont {Brahms}\ \emph {et~al.}(2012)\citenamefont {Brahms},
  \citenamefont {Botter}, \citenamefont {Schreppler}, \citenamefont {Brooks},\
  and\ \citenamefont {{Stamper-Kurn}}}]{Brahms2012}%
  \BibitemOpen
  \bibfield  {author} {\bibinfo {author} {\bibfnamefont {N.}~\bibnamefont
  {Brahms}}, \bibinfo {author} {\bibfnamefont {T.}~\bibnamefont {Botter}},
  \bibinfo {author} {\bibfnamefont {S.}~\bibnamefont {Schreppler}}, \bibinfo
  {author} {\bibfnamefont {D.~W.~C.}\ \bibnamefont {Brooks}},\ and\ \bibinfo
  {author} {\bibfnamefont {D.~M.}\ \bibnamefont {{Stamper-Kurn}}},\ }\bibfield
  {title} {\bibinfo {title} {\emph{Optical {{Detection}} of the
  {{Quantization}} of {{Collective Atomic Motion}}}},\ }\href
  {https://doi.org/10.1103/PhysRevLett.108.133601} {\bibfield  {journal}
  {\bibinfo  {journal} {Phys. Rev. Lett.}\ }\textbf {\bibinfo {volume} {108}},\
  \bibinfo {pages} {133601} (\bibinfo {year} {2012})}\BibitemShut {NoStop}%
\bibitem [{\citenamefont {Gajdacz}\ \emph {et~al.}(2016)\citenamefont
  {Gajdacz}, \citenamefont {Hilliard}, \citenamefont {Kristensen},
  \citenamefont {Pedersen}, \citenamefont {Klempt}, \citenamefont {Arlt},\ and\
  \citenamefont {Sherson}}]{Gajdacz2016}%
  \BibitemOpen
  \bibfield  {author} {\bibinfo {author} {\bibfnamefont {M.}~\bibnamefont
  {Gajdacz}}, \bibinfo {author} {\bibfnamefont {A.~J.}\ \bibnamefont
  {Hilliard}}, \bibinfo {author} {\bibfnamefont {M.~A.}\ \bibnamefont
  {Kristensen}}, \bibinfo {author} {\bibfnamefont {P.~L.}\ \bibnamefont
  {Pedersen}}, \bibinfo {author} {\bibfnamefont {C.}~\bibnamefont {Klempt}},
  \bibinfo {author} {\bibfnamefont {J.~J.}\ \bibnamefont {Arlt}},\ and\
  \bibinfo {author} {\bibfnamefont {J.~F.}\ \bibnamefont {Sherson}},\
  }\bibfield  {title} {\bibinfo {title} {\emph{Preparation of {{Ultracold Atom
  Clouds}} at the {{Shot Noise Level}}}},\ }\href
  {https://doi.org/10.1103/PhysRevLett.117.073604} {\bibfield  {journal}
  {\bibinfo  {journal} {Phys. Rev. Lett.}\ }\textbf {\bibinfo {volume} {117}},\
  \bibinfo {pages} {073604} (\bibinfo {year} {2016})}\BibitemShut {NoStop}%
\bibitem [{\citenamefont {Hadzibabic}\ and\ \citenamefont
  {Dalibard}(2011)}]{Hadzibabic2011}%
  \BibitemOpen
  \bibfield  {author} {\bibinfo {author} {\bibfnamefont {Z.}~\bibnamefont
  {Hadzibabic}}\ and\ \bibinfo {author} {\bibfnamefont {J.}~\bibnamefont
  {Dalibard}},\ }\bibfield  {title} {\bibinfo {title} {\emph{Two-dimensional
  {{Bose}} fluids: {{An}} atomic physics perspective}},\ }\href
  {https://doi.org/10.1393/ncr/i2011-10066-3} {\bibfield  {journal} {\bibinfo
  {journal} {Riv. Nuovo Cim.}\ }\textbf {\bibinfo {volume} {34}},\ \bibinfo
  {pages} {389} (\bibinfo {year} {2011})}\BibitemShut {NoStop}%
\bibitem [{\citenamefont {Periwal}\ \emph {et~al.}(2021)\citenamefont
  {Periwal}, \citenamefont {Cooper}, \citenamefont {Kunkel}, \citenamefont
  {Wienand}, \citenamefont {Davis},\ and\ \citenamefont
  {{Schleier-Smith}}}]{Periwal2021}%
  \BibitemOpen
  \bibfield  {author} {\bibinfo {author} {\bibfnamefont {A.}~\bibnamefont
  {Periwal}}, \bibinfo {author} {\bibfnamefont {E.~S.}\ \bibnamefont {Cooper}},
  \bibinfo {author} {\bibfnamefont {P.}~\bibnamefont {Kunkel}}, \bibinfo
  {author} {\bibfnamefont {J.~F.}\ \bibnamefont {Wienand}}, \bibinfo {author}
  {\bibfnamefont {E.~J.}\ \bibnamefont {Davis}},\ and\ \bibinfo {author}
  {\bibfnamefont {M.}~\bibnamefont {{Schleier-Smith}}},\ }\bibfield  {title}
  {\bibinfo {title} {\emph{Programmable {{Interactions}} and {{Emergent
  Geometry}} in an {{Atomic Array}}}},\ }\href
  {https://arxiv.org/abs/2106.04070} {\bibfield  {journal} {\bibinfo  {journal}
  {arXiv:2106.04070}\ } (\bibinfo {year} {2021})}\BibitemShut {NoStop}%
\bibitem [{\citenamefont {Schuckert}\ and\ \citenamefont
  {Knap}(2020)}]{Schuckert2020}%
  \BibitemOpen
  \bibfield  {author} {\bibinfo {author} {\bibfnamefont {A.}~\bibnamefont
  {Schuckert}}\ and\ \bibinfo {author} {\bibfnamefont {M.}~\bibnamefont
  {Knap}},\ }\bibfield  {title} {\bibinfo {title} {\emph{Probing eigenstate
  thermalization in quantum simulators via fluctuation-dissipation
  relations}},\ }\href {https://doi.org/10.1103/PhysRevResearch.2.043315}
  {\bibfield  {journal} {\bibinfo  {journal} {Phys. Rev. Research}\ }\textbf
  {\bibinfo {volume} {2}},\ \bibinfo {pages} {043315} (\bibinfo {year}
  {2020})}\BibitemShut {NoStop}%
\bibitem [{\citenamefont {Ferrier}\ \emph {et~al.}(2016)\citenamefont
  {Ferrier}, \citenamefont {Arakawa}, \citenamefont {Hata}, \citenamefont
  {Fujiwara}, \citenamefont {Delagrange}, \citenamefont {Weil}, \citenamefont
  {Deblock}, \citenamefont {Sakano}, \citenamefont {Oguri},\ and\ \citenamefont
  {Kobayashi}}]{Ferrier2016}%
  \BibitemOpen
  \bibfield  {author} {\bibinfo {author} {\bibfnamefont {M.}~\bibnamefont
  {Ferrier}}, \bibinfo {author} {\bibfnamefont {T.}~\bibnamefont {Arakawa}},
  \bibinfo {author} {\bibfnamefont {T.}~\bibnamefont {Hata}}, \bibinfo {author}
  {\bibfnamefont {R.}~\bibnamefont {Fujiwara}}, \bibinfo {author}
  {\bibfnamefont {R.}~\bibnamefont {Delagrange}}, \bibinfo {author}
  {\bibfnamefont {R.}~\bibnamefont {Weil}}, \bibinfo {author} {\bibfnamefont
  {R.}~\bibnamefont {Deblock}}, \bibinfo {author} {\bibfnamefont
  {R.}~\bibnamefont {Sakano}}, \bibinfo {author} {\bibfnamefont
  {A.}~\bibnamefont {Oguri}},\ and\ \bibinfo {author} {\bibfnamefont
  {K.}~\bibnamefont {Kobayashi}},\ }\bibfield  {title} {\bibinfo {title}
  {\emph{Universality of non-equilibrium fluctuations in strongly correlated
  quantum liquids}},\ }\href {https://doi.org/10.1038/nphys3556} {\bibfield
  {journal} {\bibinfo  {journal} {Nature Phys}\ }\textbf {\bibinfo {volume}
  {12}},\ \bibinfo {pages} {230} (\bibinfo {year} {2016})}\BibitemShut
  {NoStop}%
\bibitem [{\citenamefont {{Halpin-Healy}}\ and\ \citenamefont
  {Takeuchi}(2015)}]{Halpin-Healy2015}%
  \BibitemOpen
  \bibfield  {author} {\bibinfo {author} {\bibfnamefont {T.}~\bibnamefont
  {{Halpin-Healy}}}\ and\ \bibinfo {author} {\bibfnamefont {K.~A.}\
  \bibnamefont {Takeuchi}},\ }\bibfield  {title} {\bibinfo {title} {\emph{A
  {{KPZ Cocktail}}-{{Shaken}}, Not {{Stirred}}...}},\ }\href
  {https://doi.org/10.1007/s10955-015-1282-1} {\bibfield  {journal} {\bibinfo
  {journal} {J Stat Phys}\ }\textbf {\bibinfo {volume} {160}},\ \bibinfo
  {pages} {794} (\bibinfo {year} {2015})}\BibitemShut {NoStop}%
\bibitem [{\citenamefont {Wei}\ \emph {et~al.}(2021)\citenamefont {Wei},
  \citenamefont {{Rubio-Abadal}}, \citenamefont {Ye}, \citenamefont {Machado},
  \citenamefont {Kemp}, \citenamefont {Srakaew}, \citenamefont {Hollerith},
  \citenamefont {Rui}, \citenamefont {Gopalakrishnan}, \citenamefont {Yao},
  \citenamefont {Bloch},\ and\ \citenamefont {Zeiher}}]{Wei2021}%
  \BibitemOpen
  \bibfield  {author} {\bibinfo {author} {\bibfnamefont {D.}~\bibnamefont
  {Wei}}, \bibinfo {author} {\bibfnamefont {A.}~\bibnamefont {{Rubio-Abadal}}},
  \bibinfo {author} {\bibfnamefont {B.}~\bibnamefont {Ye}}, \bibinfo {author}
  {\bibfnamefont {F.}~\bibnamefont {Machado}}, \bibinfo {author} {\bibfnamefont
  {J.}~\bibnamefont {Kemp}}, \bibinfo {author} {\bibfnamefont {K.}~\bibnamefont
  {Srakaew}}, \bibinfo {author} {\bibfnamefont {S.}~\bibnamefont {Hollerith}},
  \bibinfo {author} {\bibfnamefont {J.}~\bibnamefont {Rui}}, \bibinfo {author}
  {\bibfnamefont {S.}~\bibnamefont {Gopalakrishnan}}, \bibinfo {author}
  {\bibfnamefont {N.~Y.}\ \bibnamefont {Yao}}, \bibinfo {author} {\bibfnamefont
  {I.}~\bibnamefont {Bloch}},\ and\ \bibinfo {author} {\bibfnamefont
  {J.}~\bibnamefont {Zeiher}},\ }\bibfield  {title} {\bibinfo {title}
  {\emph{Quantum gas microscopy of {{Kardar}}-{{Parisi}}-{{Zhang}}
  superdiffusion}},\ }\href {https://arxiv.org/abs/2107.00038} {\bibfield
  {journal} {\bibinfo  {journal} {arXiv:2107.00038}\ } (\bibinfo {year}
  {2021})}\BibitemShut {NoStop}%
\bibitem [{\citenamefont {Yang}\ \emph {et~al.}(2020)\citenamefont {Yang},
  \citenamefont {Grankin}, \citenamefont {Sieberer}, \citenamefont {Vasilyev},\
  and\ \citenamefont {Zoller}}]{Yang2020}%
  \BibitemOpen
  \bibfield  {author} {\bibinfo {author} {\bibfnamefont {D.}~\bibnamefont
  {Yang}}, \bibinfo {author} {\bibfnamefont {A.}~\bibnamefont {Grankin}},
  \bibinfo {author} {\bibfnamefont {L.~M.}\ \bibnamefont {Sieberer}}, \bibinfo
  {author} {\bibfnamefont {D.~V.}\ \bibnamefont {Vasilyev}},\ and\ \bibinfo
  {author} {\bibfnamefont {P.}~\bibnamefont {Zoller}},\ }\bibfield  {title}
  {\bibinfo {title} {\emph{Quantum non-demolition measurement of a many-body
  {{Hamiltonian}}}},\ }\href {https://doi.org/10.1038/s41467-020-14489-5}
  {\bibfield  {journal} {\bibinfo  {journal} {Nat. Commun.}\ }\textbf {\bibinfo
  {volume} {11}},\ \bibinfo {pages} {775} (\bibinfo {year} {2020})}\BibitemShut
  {NoStop}%
\bibitem [{\citenamefont {S{\"o}ding}\ \emph {et~al.}(1999)\citenamefont
  {S{\"o}ding}, \citenamefont {{Gu{\'e}ry-Odelin}}, \citenamefont {Desbiolles},
  \citenamefont {Chevy}, \citenamefont {Inamori},\ and\ \citenamefont
  {Dalibard}}]{Soding1999}%
  \BibitemOpen
  \bibfield  {author} {\bibinfo {author} {\bibfnamefont {J.}~\bibnamefont
  {S{\"o}ding}}, \bibinfo {author} {\bibfnamefont {D.}~\bibnamefont
  {{Gu{\'e}ry-Odelin}}}, \bibinfo {author} {\bibfnamefont {P.}~\bibnamefont
  {Desbiolles}}, \bibinfo {author} {\bibfnamefont {F.}~\bibnamefont {Chevy}},
  \bibinfo {author} {\bibfnamefont {H.}~\bibnamefont {Inamori}},\ and\ \bibinfo
  {author} {\bibfnamefont {J.}~\bibnamefont {Dalibard}},\ }\bibfield  {title}
  {\bibinfo {title} {\emph{Three-body decay of a rubidium
  {{Bose}}\textendash{{Einstein}} condensate}},\ }\href
  {https://doi.org/10.1007/s003400050805} {\bibfield  {journal} {\bibinfo
  {journal} {Appl Phys B}\ }\textbf {\bibinfo {volume} {69}},\ \bibinfo {pages}
  {257} (\bibinfo {year} {1999})}\BibitemShut {NoStop}%
\bibitem [{\citenamefont {Weber}\ \emph {et~al.}(2003)\citenamefont {Weber},
  \citenamefont {Herbig}, \citenamefont {Mark}, \citenamefont {N{\"a}gerl},\
  and\ \citenamefont {Grimm}}]{Weber2003}%
  \BibitemOpen
  \bibfield  {author} {\bibinfo {author} {\bibfnamefont {T.}~\bibnamefont
  {Weber}}, \bibinfo {author} {\bibfnamefont {J.}~\bibnamefont {Herbig}},
  \bibinfo {author} {\bibfnamefont {M.}~\bibnamefont {Mark}}, \bibinfo {author}
  {\bibfnamefont {H.-C.}\ \bibnamefont {N{\"a}gerl}},\ and\ \bibinfo {author}
  {\bibfnamefont {R.}~\bibnamefont {Grimm}},\ }\bibfield  {title} {\bibinfo
  {title} {\emph{Three-{{Body Recombination}} at {{Large Scattering Lengths}}
  in an {{Ultracold Atomic Gas}}}},\ }\href
  {https://doi.org/10.1103/PhysRevLett.91.123201} {\bibfield  {journal}
  {\bibinfo  {journal} {Phys. Rev. Lett.}\ }\textbf {\bibinfo {volume} {91}},\
  \bibinfo {pages} {123201} (\bibinfo {year} {2003})}\BibitemShut {NoStop}%
\bibitem [{\citenamefont {Lye}\ \emph {et~al.}(2003)\citenamefont {Lye},
  \citenamefont {Hope},\ and\ \citenamefont {Close}}]{Lye2003}%
  \BibitemOpen
  \bibfield  {author} {\bibinfo {author} {\bibfnamefont {J.~E.}\ \bibnamefont
  {Lye}}, \bibinfo {author} {\bibfnamefont {J.~J.}\ \bibnamefont {Hope}},\ and\
  \bibinfo {author} {\bibfnamefont {J.~D.}\ \bibnamefont {Close}},\ }\bibfield
  {title} {\bibinfo {title} {\emph{Nondestructive dynamic detectors for
  {{Bose}}-{{Einstein}} condensates}},\ }\href
  {https://doi.org/10.1103/PhysRevA.67.043609} {\bibfield  {journal} {\bibinfo
  {journal} {Phys. Rev. A}\ }\textbf {\bibinfo {volume} {67}},\ \bibinfo
  {pages} {043609} (\bibinfo {year} {2003})}\BibitemShut {NoStop}%
\bibitem [{\citenamefont {Hume}\ \emph {et~al.}(2013)\citenamefont {Hume},
  \citenamefont {Stroescu}, \citenamefont {Joos}, \citenamefont {Muessel},
  \citenamefont {Strobel},\ and\ \citenamefont {Oberthaler}}]{Hume2013}%
  \BibitemOpen
  \bibfield  {author} {\bibinfo {author} {\bibfnamefont {D.~B.}\ \bibnamefont
  {Hume}}, \bibinfo {author} {\bibfnamefont {I.}~\bibnamefont {Stroescu}},
  \bibinfo {author} {\bibfnamefont {M.}~\bibnamefont {Joos}}, \bibinfo {author}
  {\bibfnamefont {W.}~\bibnamefont {Muessel}}, \bibinfo {author} {\bibfnamefont
  {H.}~\bibnamefont {Strobel}},\ and\ \bibinfo {author} {\bibfnamefont {M.~K.}\
  \bibnamefont {Oberthaler}},\ }\bibfield  {title} {\bibinfo {title}
  {\emph{Accurate {{Atom Counting}} in {{Mesoscopic Ensembles}}}},\ }\href
  {https://doi.org/10.1103/PhysRevLett.111.253001} {\bibfield  {journal}
  {\bibinfo  {journal} {Phys. Rev. Lett.}\ }\textbf {\bibinfo {volume} {111}},\
  \bibinfo {pages} {253001} (\bibinfo {year} {2013})}\BibitemShut {NoStop}%
\end{thebibliography}%

\clearpage

%\section*{Appendix}
\section*{Appendix A: Methods}
\subsection*{Experimental sequence}
\begin{figure}
  \centering
  \includegraphics{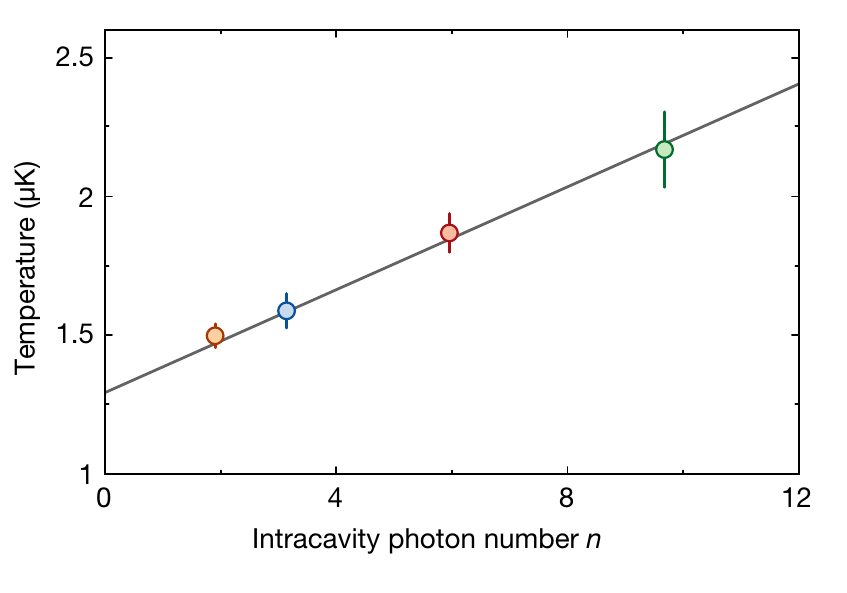}%
  \caption{\label{fig:S4}
  \textbf{Temperature after evaporation.}
  Temperature of the gas measured in time of flight after the final swept atom number measurement for different intracavity photon numbers. The gray line is a linear fit guiding the eye.}
\end{figure}

We started our measurement by preparing a cloud of about $2200$ $^\mathrm{87}$Rb atoms with a temperature of approximately $T_0=2.6\,\mu$K in a cavity-supported vertical optical standing wave, forming an optical lattice with lattice spacing $a = 421\,$nm. Before loading the lattice, the cloud was compressed vertically in a tight magnetic trap created by an atom chip, such that predominantly a single slice of the lattice is populated~\cite{Purdy2010a,Kohler2017,Kohler2018}. We estimate the fraction of atoms in other slices to be maximally $20\%$, as was characterized by cavity-aided magnetic resonance microscopy~\cite{Brahms2011}. The depth of the lattice trap was $U_0/2\pi=h\times641(30)\,$kHz ($k_B\times31(1)\,\mu$K) and the waist of the lattice beam in the cavity was $w_0=26\,\mu$m, leading to axial and radial trapping frequencies of $\omega_z/2\pi = 91(2)\,$kHz and $\omega_r/2\pi=670(10)\,$Hz. Given $k_BT_0=h\times54\,$kHz, these parameters put the gas in a quasi two-dimensional regime, where motion along the $z$-direction is frozen out, but atomic collisions are still described by a three-dimensional scattering process. The cavity length and exact wavelength of the lattice laser were chosen such that the atomic cloud axially overlapped with the maximal probe intensity in the cavity.
The atoms emerged from the magnetic trap purely in the $\ket{F,\,m_F}=\ket{2,\,2}$ hyperfine state.  During probing, we applied a strong magnetic field of $B_z=17\,$G along the cavity axis, and also drove the cavity with $\sigma^+$ circularly polarized probe light.  Under these conditions, the light-atom interactions are reduced to an effective two-level scheme, the atomic spin polarization is preserved, and the vacuum Rabi coupling is maximized.\\
The atom number was recorded using a cavity probe at a wavelength of $780\,$nm, detuned by $\Delta_{ca}/2\pi\approx-42\,$GHz to the red of the $D2$ line.
The transmission of this probe through the cavity was recorded on a heterodyne receiver.
For heterodyne detection, we used a local oscillator (LO) with approximately $1\,$mW of optical power derived from the same laser but with its frequency offset by $10\,$MHz relative to the frequency of the probe. The probe beam was intensity stabilized before entering the cavity and the LO before coupling into the heterodyne receiver. The photon collection efficiency was reduced by cavity mirror losses ($\epsilon_c = 0.31$), finite path efficiency of the heterodyne path ($\epsilon_p = 0.93$), mode matching efficiency of local oscillator and probe beam ($\epsilon_{mm} =0.89$) and quantum efficiency of the heterodyne photodetector ($\epsilon_q \approx 0.58$). These effects combined to give an overall photon detection efficiency of $\epsilon = 0.149(10)$ in our heterodyne receiver.\\
\begin{figure}
  \centering
  \includegraphics{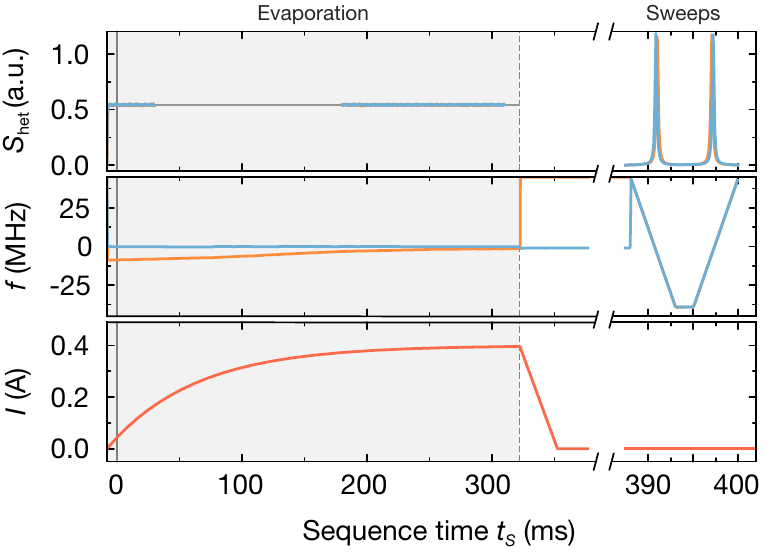}%
  \caption{\label{fig:S1}
  \textbf{Experimental sequence.}
  Upper panel: Heterodyne magnitude $S_\mathrm{het}$ with atoms present (not present) in orange (blue). The heterodyne magnitude signal was not recorded for times $t_s$ between $30\,$ms and $170\,$ms. Middle panel: Corresponding VCO frequency $f$. Bottom panel: Current in our atomic chip wires (red). Evaporation was induced by ramping up the current, which lowers the trap depth dynamically (see Fig.~\ref{fig:S2}), and halted at $320\,$ms by ramping the current back down quickly. During evaporation (indicated by gray shading in all panels), the heterodyne magnitude was kept constant by a side-of-fringe lock. The atom-induced cavity shift was extracted from an in-loop measurement of the control voltage fed back to the VCO. At the end of an experimental run, a swept atom number measurement was performed, where $\delta_{pc}$ was varied by sweeping $f$ down and up across cavity resonance. In these sweep measurements, the atom number was extracted from the shift of the cavity profile visible in the heterodyne magnitude signal, and was used just for calibration purposes.}
\end{figure}
After preparation of the cloud, forced evaporation was induced by exponentially ramping up the current through our atomic chip wires with a time constant of $70\,$ms. The resulting inhomogeneous magnetic field was accompanied by a magnetic field gradient increasing from zero up to $220\,\mathrm{G}/\mathrm{cm}$, strong enough to lower the trap depth dynamically; see Fig.~\ref{fig:S2}.
During the evaporation process, an analog side-of-fringe feedback loop was engaged to keep the probe-cavity detuning fixed at $\delta_{pc}/2\pi\approx\kappa/2\pi=1.8\,$MHz and thus the intracavity probe intensity constant. The time-zero of our measurement was chosen $8\,$ms after activating the feedback, which was enough time to avoid any effect of transients on our real-time atom-number traces. We believe, however, that these transients and variations caused by loading the cloud in our cavity lattice contribute to the initial temperature variation observed in our gas.\\ 
To realize the side-of-fringe feedback loop, part of the heterodyne signal was split off after a radio-frequency amplifier, sent through a bandpass filter with a $2\,$MHz bandwidth, and its power was detected with a linear radio-frequency power detector (Analog Devices, AD$8361$). The detected power was kept constant by feeding back to the frequency of probe and LO through a voltage controlled oscillator (VCO) driving an acousto-optic modulator (AO) before the cavity. The required in-loop control voltage was monitored. An equivalent VCO signal was recorded after the sequence without the atoms present. The difference of this reference signal and the signal with atoms, together with the calibrated VCO characteristics, yielded the atom-induced cavity shift $\Delta_N$. In order to reduce the effect of photon shot noise, we subtracted a running average of five empty cavity traces of successive runs from each trace with atoms. The number of averaged empty cavity traces was a compromise between mitigating the influence of shot noise on the reference trace and avoiding additional imprecision due to longer-time technical drifts of the empty cavity resonance.
After untilting the trap, we performed a swept atom number measurement; see Fig.~\ref{fig:S1}. To this end, we swept the frequency of the cavity probe across the resonance of our cavity. The peak position of the recorded heterodyne signal relative to a reference measurement on an empty cavity taken at the end of the sequence reflected the atom-induced cavity shift; see Fig.~\ref{fig:S1}.
The swept measurement was used to benchmark the shift correction described in the next section.\\
In order to verify the cooling of our gas, we performed a time-of-flight absorption measurement after our swept atom number measurement. To this end, we rapidly turned off the cavity lattice and let the cloud expand by $400\,\mu$s. The temperatures extracted from this time of flight measurement are shown in Fig.~\ref{fig:S4}. We consider this extracted temperature to be an upper bound of the actual temperature due to the long additional hold time before the measurement, the additionally present swept atom number measurement, and the short expansion time of the gas limited by our cavity geometry. 
We note that the temperatures extracted from our simulation right at the end of the evaporation ramp are approximately a factor of two below these values; see Fig.~\ref{fig:S3}.
\subsection*{Trap depth and shift correction}
\begin{figure}
  \centering
  \includegraphics{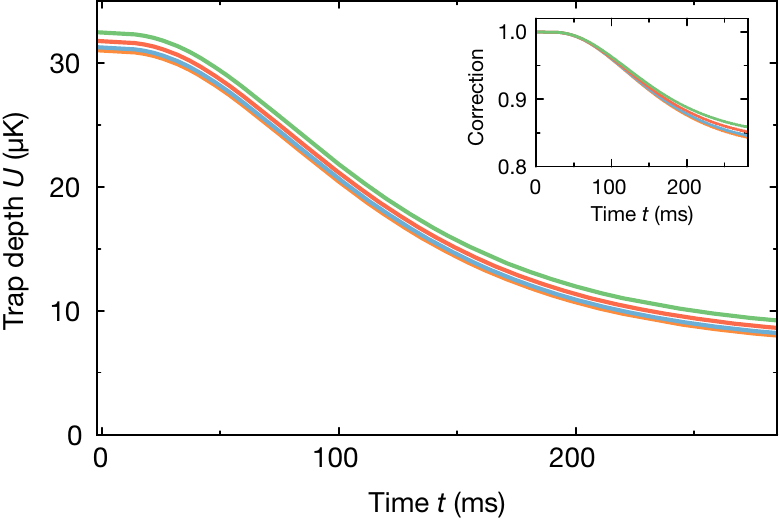}%
  \caption{\label{fig:S2}
  \textbf{Potential depth during forced evaporation.}
  The trap depth of the combined potential of far off-resonant optical lattice trap, magnetic field and probe-induced potential as the magnetic gradient is ramped up from zero to its maximal value. The inset shows the correction factor required to correct $g^2$ for the shift of the cloud relative to the maximal probe intensity. Different colors correspond to different intracavity photon number, with the color code the same as in Fig.~\ref{fig:2}.}
\end{figure}

The magnetic field applied during evaporative cooling leads to a reduction of the trap depth to below $10\,\mu$K; see Fig.~\ref{fig:S2}. We find the trap depth using the known waist size of the lattice trap in the cavity and the applied magnetic fields as the height of the saddle point of the combined potential of magnetic and optical traps relative to the trap bottom, including the small but non-negligible influence of the red-detuned probe. With increasing applied magnetic field gradients, the trap minimum is radially displaced from the center of the cavity mode, and hence away from the maximal coupling point of the probe laser. We calculate the displacement and the resulting reduced overlap with the probe laser for a point-like cloud. The maximal displacement is approximately $8\,\mu$m. For our probe waist of $w_p = 25\,\mu$m, this displacement leads to a change in $g^2$ and hence $\Delta_{1}$ of at most $15\,\%$ (see the inset of Fig.~\ref{fig:S2}). We verified that the results of this procedure for a point-like cloud were almost identical to the case of a thermal cloud with a temperature of $T_0\approx2.6\,\mu$K, such that we use the former as an approximation for the latter. We confirm the accuracy of our calculated correction factor by comparing the corrected atom number measured at the maximal spatial displacement with a sweep measurement after removing the magnetic field gradient, see Fig.~\ref{fig:S1}, where the atoms were located at the position of maximal probe coupling. Due to the strong lattice-induced confinement, the displacement of the cloud along the vertical direction is negligible. In addition to the sideways displacement, the applied magnetic field rotates the overall magnetic field direction. Consequently, the orientation of the quantization axis tilts and the projection of the circularly polarized light of the probe beam changes, resulting in maximally $3\%$ of additional correction to the coupling strength. We estimate an associated maximal probability of $4\%$ per scattered photon to result in transfer into other hyperfine ground states, with a negligible effect on our measured atom numbers.
We also stress that the small correction of $g^2$ does not affect any of the conclusions drawn, and we have accounted for it in the direct quantitative comparison of our measurements with the theoretical models presented in the following.
\section*{Appendix B: Modeling evaporative cooling}
\subsection*{Theoretical model for evaporation}\label{sec:Theomodel}
We model evaporative cooling in two dimensions, following references~\cite{Luiten1996a,OHara2001a}, by the time evolution of a truncated Maxwell-Boltzmann distribution in a harmonic trap with time-dependent depth.
The change in atom number $N$ for a truncation parameter $\eta=U/k_BT$ in the trap is
\begin{align}\label{eq:SI_Ndot}
\dot{N}=-N\gamma e^{-\eta}\left(\frac{\eta P(2,\eta)-3P(3,\eta)}{P(2,\eta)^2}\right).
\end{align}
Here, $\gamma=n_{th}\sigma v_{th}$ is related to the elastic collision rate $\gamma_c = \sqrt{2}\gamma$ in a gas with density $n_{th}$ and thermal velocity $v_{th} = \sqrt{8k_BT/\pi m}$. The incomplete Gamma functions $P(\alpha,\eta)=\int_0^{\eta}\mathrm{d}t\,t^{\alpha-1}e^{-t}/\Gamma(\alpha)$ take into account truncation effects in a trap with finite depth~\cite{Luiten1996a}.
We can rewrite Eq.~\eqref{eq:SI_Ndot} to bring out the temperature and atom number dependence explicitly,
\begin{align}\label{eq:SI_Ndot}
\dot{N}=-c_\gamma N^2T^{-\frac{1}{2}}e^{-\eta}\left(\frac{\eta P(2,\eta)-3P(3,\eta)}{P(2,\eta)^2}\right).
\end{align}
The constant $c_\gamma=\gamma_0T_0^{\frac{1}{2}}/N_0$ contains the initial rate $\gamma_0$, the initial temperature $T_0$ and the initial atom number $N_0$. Notably, Eq.~\eqref{eq:SI_Ndot} indicates that, at constant temperature, atom spilling over a potential barrier constitutes a loss process which is non-linear in atom number.\\
Evaporating atoms cause a change in the internal energy $E=2Nk_BT$ of the gas. Taking the time derivative and solving for the change in temperature, we obtain
\begin{align}\label{eq:SI_Tdot1}
k_B\dot{T} = \frac{\dot{E}}{2N}-\frac{\dot{N}}{N}k_BT.
\end{align}
The change in energy is proportional to the atom loss rate multiplied with the energy of a leaving atom, corrected for truncation~\cite{Luiten1996a} 
\begin{align}
\dot{E}=\dot{N}k_BT\left[\eta+\left(1-\frac{P(4,\eta)}{\eta P(2,\eta)-3P(3,\eta)}\right)\right],
\end{align}
where $\eta = U/k_BT$.
Plugging this into Eq.~\eqref{eq:SI_Tdot1}, we obtain the change in temperature as
\begin{align}\label{eq:SI_Tdot2}
\frac{\dot{T}}{T}=\frac{1}{2}\frac{\dot{N}}{N}\left[\eta-\left(1+\frac{P(4,\eta)}{\eta P(2,\eta)-3P(3,\eta)}\right)\right]+\frac{\Gamma_h}{2T}.
\end{align}
In the last step, we have included a heating rate $\Gamma_h$, caused, e.g., by photon recoil heating or other external heating sources.\\
Three-body losses affect both the atom number dynamics and the temperature dynamics of evaporative cooling. Their atom number dynamics have to be treated separately from the atom number dynamics due to evaporating atoms. The atom loss rate due to three-body collisions is
\begin{align}\label{eq:threebodyNdot}
\dot{N}_{3B} = -c_3\frac{6}{\sqrt{27}}L_3\,n_{th}^2\,N \propto \frac{N^3}{T^2}.
\end{align}
Here, $L_3$ denotes the three-body loss coefficient~\cite{Soding1999}, which is scaled appropriately for a thermal bosonic gas~\cite{Weber2003}, and $c_3$ is an additional scaling coefficient.
These losses do not contribute to the cooling of the gas. Rather, they lead to ``anti-evaporation heating''~\cite{Weber2003}, as the loss happens predominantly at the center of the trap, where the density is highest. Atoms in the central region of the trap have less than average potential energy, such that the gas effectively heats upon thermalization.
Quantitatively, anti-evaporation heating leads to a change in temperature reflecting the difference in potential energy between an average atom and an atom lost through a three-body collision,
\begin{align}\label{eq:SI_threebodyTdot}
\left(\frac{\dot{T}}{T}\right)_{3B} = -\frac{1}{2}\frac{\dot{N}_{3B}}{N}\left(1-\sqrt{\frac{2}{3}}\right)
\end{align}
Together with the known time-dependent trajectory of the trap depth $U(t)$, equations~\eqref{eq:SI_Ndot} and~\eqref{eq:SI_Tdot2} can be solved to obtain the mean atom number and temperature dynamics of evaporative cooling. The effect of three-body loss can be included by adding equations~\eqref{eq:threebodyNdot} to~\eqref{eq:SI_Ndot} and~\eqref{eq:SI_threebodyTdot} to~\eqref{eq:SI_Tdot2}. 
\subsection*{Evaporation dynamics}
\begin{figure*}
  \centering
  \includegraphics[width =\textwidth]{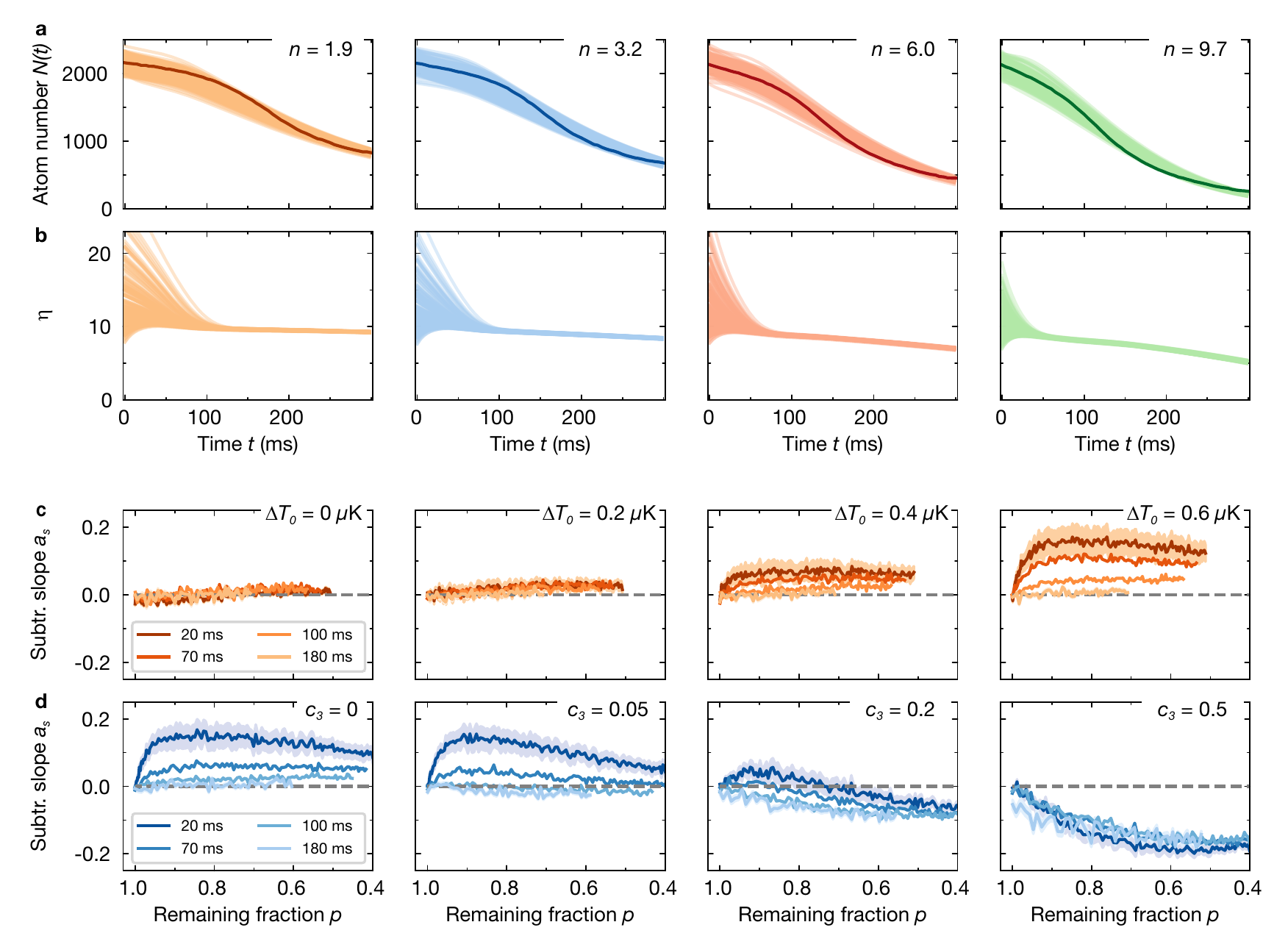}%
  \caption{\label{fig:S3}
  \textbf{Theoretical modeling of evaporation.}
  \textbf{a} Simulation of atom number dynamics during forced evaporation for the trap ramp shown in Fig.~\ref{fig:S2}. We show $200$ traces with initial atom numbers $N_0$ and temperatures $T_0$ sampled from Gaussian distributions in light color, for varying intracavity photon number $n$ as indicated in the top right corner. The atom number samples have a mean $\avg{N_0}$ and a standard deviation of $2\avg{N_0}^{1/2}$. The temperature distribution is centered about the initial temperature $T_0$ and has a standard deviation of $\Delta T_0 = 0.6\,\mu$K. We have included a three-body loss contribution scaled by $c_3 = 0.2$. The dark line shows the measured atom number dynamics including shift correction.
  \textbf{b} Ratio of trap depth to temperature, $\eta = U/k_BT$, for the same ensemble as in \textbf{a}. The initially broad distribution is rapidly narrowed down and then remains nearly constant during further evaporation.
  \textbf{c} Subtracted slope $a_s$ for $n=1.9$ for different temperature spreads (indicated in top right corner) in absence of three-body loss ($c_3=0$). We observe the same trends as observed in our experiment, with a fast rise for small $p$ and a decay with increasing initial time $t_1$ (indicated in legend). The shaded region indicates the standard deviation extracted from a bootstrap analysis of the calculated samples.
  \textbf{d} Subtracted slope $a_s$ for $n=3.2$ and $\Delta T_0 = 0.6\,\mu$K for increasing contribution of three-body loss characterized by the scaling factor $c_3$ (indicated in top right corner). Three-body loss leads to a negative contribution to $a_s$, which suppresses the effect of initial temperature variation.}
\end{figure*}
The model derived in the previous section can be used to check the influence of initial atom number and temperature variation on the evaporation dynamics, and serves as a benchmark of the subtracted slope presented in the main text as a measure of non-linear dynamics in the average evolution.
To this end, we simulate the coupled atom- and temperature dynamics for our experimental configuration. We neglect the influence of adiabatic decompression~\cite{OHara2001a}, which is expected to be small in gradient-assisted evaporation~\cite{Hung2008a}. In our case, the expected reduction of the trap frequency along the direction of the gradient is approximately $15\%$.
The heating rate per photon is extracted from reference measurements to be $\Gamma_{h}=7(3)\,\mu\mathrm{K\,s}^{-1}$. For our parameters, we expect scattering into free space to be a significant heating source. The off-resonant scattering rate $\Gamma_\mathrm{eff}/n\approx 3.6\,\mathrm{s}^{-1}$ leads to an expected recoil heating rate per intracavity photon of $\Gamma_r/n = 2E_r\,\Gamma_\mathrm{eff}/n \approx h\times27.1\,\mathrm{kHz}\,\mathrm{s}^{-1}=k_B\times1.3\,\mu\mathrm{K\,s}^{-1}$, where we have used the recoil energy $E_r = \hbar^2k^2/2m \approx h\times3.8\,$kHz for $^\mathrm{87}$Rb on the $D2$ line. We attribute the discrepancy between this estimate and the experimental value to additional technical noise contributions from the side-of-fringe lock, which can lead to parametric heating of the cloud.\\
To model variations in the initially prepared ensemble, we calculate $200$ traces with atom number $N_0$ and temperature $T_0$ drawn from Gaussian distributions. The standard deviation of the atom number distribution is extracted from a fit to the measured initial atom number distribution. The standard deviation of the temperature distribution $\Delta T_0$ is varied in the simulation. The elastic collision rate is calculated for each run using the drawn atom number and temperature assuming a quasi two-dimensional gas.
The resulting curves of the time evolution of the atom number for our trap ramp are shown in Fig.~\ref{fig:S3}a, for a temperature spread of $\Delta T_0 = 0.6\,\mu$K and a three-body collision scale factor $c_3=0.2$. The overall temporal dynamics and the final atom numbers are in good agreement with the data for the chosen parameters.
During the initial $100\,$ms of the evaporation process, the temperature variation, reflected in variation in the ratio of trap depth to temperature $\eta$, are rapidly suppressed and temperature locks to an approximately fixed fraction of trap depth provided the heating is small; see Fig.~\ref{fig:S3}b.\\
In order to support the introduction of the subtracted slope $a_s = \rho_{12}\,\sigma_2/\sigma_1-p$ as a non-linearity measure (see main text and Appendix C), we study the effect of varying temperature spread and three-body loss on $a_s$. The resulting dependencies are shown in Fig.~\ref{fig:S3}c and Fig.~\ref{fig:S3}d. We observe that, as expected and explained in the main text, an increase in the initial temperature spread is reflected as a positive signal in $a_s$, which reduces to near zero and stays constant for later initial time $t_1$.
Contrarily, the introduction of strong three-body loss in the dynamics results in a negative $a_s$ and hence sub-linear behavior, as expected for an effect that reduces the atom number variance in time~\cite{Whitlock2010}, even if strong initial temperature variations are present in the ensemble; see Fig.~\ref{fig:S3}d. Qualitatively, the simulation shows the same features as observed in our data for three-body losses of approximately $c_3=0.05-0.2$ times the value expected according to Eq.~\eqref{eq:threebodyNdot}. This reduction of three-body losses points to an overestimate of the density of the ensemble, for example due to interactions in the two-dimensional gas, which we have neglected in our treatment.
Our simulations support our claim that the subtracted slope $a_s$ is a meaningful quantity sensitive to both initial temperature variation and three-body loss. We have verified that this is robustly the case also if the thermalization and heating rates are reduced. For stronger heating, we find experimentally and from our simulations that evaporation assumes linear character with $a_s=0$. 
While our simulations do reflect the effect of variation in the initial ensemble, we stress that the intrinsic stochastic character of evaporation is not captured. Further studies of the interplay between stochastic fluctuations and non-linearities such as three-body loss would be of particular interest.
\section*{Appendix C: Evaporation non-linearity and noise}\label{sec:Noise}
\subsection*{Non-linearity measure}
The minimally invasive measurement strategy involving the cavity allows to go beyond single-time observables by constructing two-time correlations.
The information about the linearity of the time evolution is encoded in the dependence of the measured atom number $N_2$ on the initial atom number $N_1$. Here, we denote $N_1$ and $N_2$ as ensembles of measured atom number outcomes at times $t_1$ and $t_2$ respectively. The standard deviations of the distributions are denoted $\sigma_1$ and $\sigma_2$ respectively.
Assuming a linear relationship between $N_1$ and $N_2$, the optimal set of coefficients $a$ and $b$ of the model $f(X,\beta) =  a X+b$ can be found from linear regression analysis for the variables $N_1$ and $N_2$.
The parameters minimizing the squared residuals are 
\begin{align}\label{eq:a}
a &= \rho_{12}\,\frac{\sigma_2}{\sigma_1}\\
b &= \avg{N_2}-a\avg{N_1}.
\end{align}
Here, we have introduced the Pearson correlation coefficient
\begin{align}\label{eq:corrcoeff}
\rho_{12}=\frac{\mathrm{cov}(N_1,N_2)}{\sigma_1\sigma_2}
\end{align}
calculated from the covariance of the two random variables. 
When standardizing the random variables by subtracting their mean and normalizing to the standard deviation, $N_{1(2),s} = (N_{1(2)}-\avg{N_{1(2)}})/\sigma_{1(2)}$, the correlation coefficient corresponds directly to the slope of the straight line minimizing the ordinary least squares in the scatter plot. A decreasing correlation coefficient in this case directly reflects the influence of measurement and stochastic noise in the system. Scatter plots of $N_{2,s}$ vs.~$N_{1,s}$ are shown in Fig.~\ref{fig:2} in the main text.
To extract the non-linearity measure, we consider non-standardized variables.
Here, for a strictly linear relationship, the fraction of remaining atoms $p=\avg{N_2}/\avg{N_1}$ contains the full information on the dynamics. Therefore, it should hold that $a=p$ in this case. Any deviation from $a=p$ can be interpreted as a non-linearity and leads to increased ($a>p$) or decreased ($a<p$) relative variation in the ensemble at time $t_2$. On average, $a<p$ ($a>p$) implies a better (worse) predictive power of a measurement result obtained at time $t_1$ for a measurement performed at time $t_2$. As an example, if the atom loss dynamics has a character $\dot{N}\propto -N^{\alpha}$, $\alpha>1$, we expect realizations with more atoms initially to lose atoms more quickly. This leads to a value of $a<p$ for later times, as more atoms were lost than expected for a linear process. Therefore, a loss process with $\alpha>1$ reduces atom number variation in an ensemble, which has been observed experimentally for a cloud with three-body loss ($\alpha = 3$)~\cite{Whitlock2010} and is consistent with the simulations shown in Fig.~\ref{fig:S3}d.
\subsection*{Unexplained variance}
The correlation coefficient quantifies the degree of fluctuation between two random variables which can be explained by a linear model.
For stochastic processes such as evaporative cooling or other transport phenomena, the fluctuations due to this randomness are also of interest.
Relating measurements of an atom number $N_2$ at time $t_2$ to measurements $N_1$ at time $t_1$ of the same ensemble can shed light on this process noise, as it appears as fluctuations in $N_2$ which are not explained by any correlation with $N_1$.\\
We can quantify this ``unexplained variance'' by considering the prediction for $N_2$ based on the value of $N_1$, which is
\begin{align}
 N_{2,pr} = \avg{N_2}+a\left(N_1-\avg{N_1}\right)
\end{align} 
This leads to the following unexplained variance
\begin{align}
\sigma_{u}^2 =\avg{\left(N_2-N_{2,pr}\right)^2}=\sigma_2^2+a^2\sigma_1^2-2a\,\mathrm{cov}(N_1,N_2)
\end{align}
This quantity is minimized for $a=\rho_{12}\,\sigma_2/\sigma_1$ (see Eq.~\eqref{eq:a}) and, thus,
\begin{align}\label{eq:varunexplained}
\sigma_u^2 = \sigma_2^2\left(1-\rho_{12}^2\right).
\end{align}
This expression shows that the correlation coefficient captures the fraction of the total variance explained by the linear relationship, which is subtracted from the total variance to obtain the unexplained variance. We evaluate the fraction of unexplained variance for our data; see Fig.~\ref{fig:4}.
\subsection*{Unexplained variance for uncorrelated atom loss}
In the following, we will derive an expression for $\sigma_u$ assuming a given constant measurement noise $\sigma_m$ and a purely Poissonian ejection mechanism of atoms with corresponding stochastic noise.
Under these assumptions, we expect the variance $\sigma_{2,th}$ at time $t_2$ to contain contributions from measurement noise, stochastic noise and variation $\sigma_1$ of $N_1$. Concretely, we get
\begin{align}\label{eq:var2theo}
\sigma_{2,th}^2 = p^2\left(\sigma_1^2-\sigma_m^2\right) + \sigma_m^2 + \avg{N_1}\,p(1-p).
\end{align} 
The second and third term are measurement noise and Poissonian stochastic noise, respectively. The first term reflects the scaled initial ensemble variation, which is obtained from the part of the variation $\sigma_1$ which is not originating from measurement noise.
The prediction for the correlation coefficient defined in Eq.~\eqref{eq:corrcoeff} becomes
\begin{align}\label{eq:corrtheo}
\rho_{12,th} = \frac{p\,\left( \sigma_1^2-\sigma_m^2 \right)}{\sqrt{\sigma_{2,th}\sigma_1}}.
\end{align}
In the covariance in the numerator, only those parts of the variances without the measurement noise contribute. Note that this is strictly true only outside the correlation time of the measurement noise, which we assume to be white for simplicity.
For equal-time measurements, the correlation coefficient equals unity.
Using Eq.~\eqref{eq:var2theo} and Eq.~\eqref{eq:corrtheo}, we can get a prediction for the fraction of variance unexplained,
\begin{align}\label{eq:sigma_u}
\sigma^2_{u,th}&=\sigma_{2,th}^2\left(1-\rho_{12,th}^2\right)\\
			   &=p^2\frac{\sigma_m^2}{\sigma_1^2} \left(\sigma_1^2-\sigma_m^2\right) + \sigma_m^2 + \avg{N_1}\,p(1-p).
\end{align}
This expression shows that the unexplained variance is affected by the initial variance, measurement noise and stochastic noise. Comparing with the expression for the variance $\sigma^2_{2,th}$, we see that the contribution from the initial variance is suppressed for small measurement noise $\sigma_m^2\ll\sigma_1^2$. For $p\rightarrow1$, the first term approaches $\sigma_m^2$, such that the total unexplained variance in this case approaches $2\sigma_m^2$.
We compare this model to our data in Fig.~\ref{fig:4}. The measurement noise $\sigma_m$ is extracted from the Allan deviation shown in Fig.~\ref{fig:1}e and the initial mean $\avg{N_1}$ and standard deviation $\sigma_1$ are evaluated from the atom number distribution at time $t_1$.  The most striking discrepancy is caused by temperature variations: Initially hidden, they are converted to strong atom number variations upon cutting into the thermal distribution. These variations are not expected based on earlier atom number measurements alone. At later times, the unexplained variance approaches the prediction for uncorrelated atom loss. The good agreement between this simple model and our data at late times indicates that technical noise has a negligible impact on our measurements. The convex shape proportional to $p(1-p)$ of fraction of variance unexplained is masked to a large part by the first term in $\sigma_{u,th}$, despite the small imprecision reached by cavity-assisted atom number detection. 

\section*{Appendix D: Measurement precision}\label{sec:Measprecision}
In the following, we show that the enhanced atom light-coupling in an optical cavity is fundamental to high-precision atom counting. Minimally invasive measurements compromise between excess heating and consecutive fluctuations due to the measurement process and a sufficient number of extracted photons to reduce photonic shot noise~\cite{Lye2003,Hume2013,Gajdacz2013}. For cavity-enhanced atom counting, we aim to find an optimal observation time as a compromise between averaging down photon shot noise of the detected probe field and probe-induced atom number fluctuations, e.g. by recoil heating or cavity backaction heating~\cite{Brahms2011,Zhang2012,Chen2014,Norcia2016}.
To quantify the photonic shot noise contribution, we use the Lorentzian cavity profile for the photon number $n$ in the cavity
\begin{align}\label{eq:1}
n = n_\mathrm{max}\frac{\kappa^2}{\kappa^2+\delta_{pc}^2}.
\end{align}
Here, $n_\mathrm{max}$ denotes the on-resonance ($\delta_{pc}=0$) intracavity photon number and $\kappa$ is the half linewidth of the cavity.
The sensitivity $\Delta\delta_{pc}$ for determining the cavity-probe detuning as a function of photon noise $\Delta n$ is obtained by taking the derivative of Eq.~\eqref{eq:1} with respect to $\delta_{pc}$,
\begin{align}
\frac{\Delta n}{n}=-2~\frac{\delta_{pc}^2}{\kappa^2+\delta_{pc}^2}~\frac{\Delta \delta_{pc}}{\delta_{pc}}.
\end{align} 
At the side of fringe for $\delta_{pc}=\kappa$, the sensitivity becomes $\frac{\Delta n}{n}=-\frac{\Delta\delta_{pc}}{\kappa}$.
Taking into account the detection efficiency $\epsilon$ of the detection chain, the number of detected photons within a window of integration time $\tau$ is $n_{\mathrm{det}}=2\kappa n \epsilon \tau$. Note that for a heterodyne detector, the detection efficiency effectively reduces by a factor $2$ ($\epsilon\rightarrow\epsilon/2$) if only the magnitude of the heterodyne signal is used in the detection, as is the case in our, in this sense, non-optimal feedback scheme.
Assuming photon-shot noise $\Delta n_\mathrm{det} = \sqrt{n_\mathrm{det}}$, we find the uncertainty
\begin{align}
\Delta\delta_{pc}=-\sqrt{\frac{\kappa}{2n\epsilon\tau}}.
\end{align}
Using the cavity shift per atom, $\Delta_{1}$, we convert this to an atom number uncertainty
\begin{align}\label{eq:2}
\Delta N = \frac{\Delta_{ca}}{g^2}\abs{\Delta\delta_{pc}} = \frac{\Delta_{ca}}{g^2}\sqrt{\frac{\kappa}{2n\epsilon\tau}} = \sqrt{\frac{1}{2C}}\sqrt{\frac{1}{\Gamma_{\mathrm{eff}}}}\sqrt{\frac{1}{\epsilon\tau}}.  
\end{align}
In the last step, we have introduced the cooperativity $C = g^2/\kappa\Gamma$ for a vacuum Rabi coupling $g=\sqrt{d^2\omega_a/2\hbar\epsilon_0V_m}$ on the optical transition with frequency $\omega_a$ and dipole matrix element $d$ and for a cavity mode volume $V_m$.
The dependence of the uncertainty on the cooperativity shows the advantage of a cavity-assisted measurement over its free-space equivalent. Furthermore, the fluctuations due to photonic shot noise decrease with the number of scattered photons $\Gamma_{\mathrm{eff}}\tau$.
The total number of scattered photons is also responsible for heating the cloud via recoil heating with a rate $\Gamma_r$, which introduces atom number loss and associated fluctuations in a trap with finite depth. Assuming Poissonian fluctuations in the number of lost atoms within an integration time $\tau$, we model the fluctuations in atom number in a trap with depth $U$ as
\begin{align}
\Delta N^2 =  N \frac{\alpha\Gamma_r}{U}\tau =N \Gamma_\mathrm{eff} \frac{\alpha E_r}{U}\tau.
\end{align}  
The proportionality constant $\alpha$ allows to take into account truncation in evaporative cooling. 
We can add the fluctuations due to atom loss to the photonic shot-noise-induced fluctuations derived in Eq.~\eqref{eq:2} and obtain
\begin{align}
\Delta N_\mathrm{tot}^2 = \frac{1}{2C}\frac{1}{\Gamma_\mathrm{eff}}\frac{1}{\epsilon\tau} +N \Gamma_\mathrm{eff} \frac{\alpha E_r}{U}\tau .
\end{align}
Minimizing the uncertainty with respect to $\tau$, we get
\begin{align}
\Delta N_\mathrm{tot,min}^2 = N\sqrt{\frac{2}{NC\epsilon}}\,\sqrt{\frac{\alpha E_r}{U}}
\end{align}
at a corresponding integration time
\begin{align}
\tau_\mathrm{min} = \frac{1}{\Gamma_\mathrm{eff}}\sqrt{\frac{1}{2NC\epsilon}}\,\sqrt{\frac{U}{E_r\alpha}}.
\end{align}
We want to minimize both the minimal fluctuations as well as the corresponding integration time. Short integration times correspond to maximal dynamic range for measuring atom number dynamics.
We see that both quantities benefit from a cavity with high cooperativity. Also, the dependence on collective cooperativity $NC$ shows that precise cavity-enhanced atom counting is easier to achieve at larger atom numbers. Interestingly, the minimal atom number imprecision does not depend on the effective scattering rate $\Gamma_\mathrm{eff}$. We observe that the minimal atom number imprecision has no fundamental limit, as the ratio $E_r/U$ can be reduced by increasing the trap depth, which reduces the loss.
If temperature is a relevant parameter in the probed dynamics, a further interesting quantity to take into account is the temperature increase of the gas during the integration time. From the increase in internal energy and assuming equipartition in a two-dimensional harmonically trapped gas, it is given as
\begin{align}
k_B\Delta T_\mathrm{min} &= \frac{2\Gamma_\mathrm{eff}E_r\tau_\mathrm{min}}{2}\\
						 &= \sqrt{\frac{1}{2NC\epsilon}}\,\sqrt{\frac{U E_r}{\alpha}}.
\end{align}
Taking the product of temperature added and minimal atom number uncertainty, the experiment-specific quantities $U$ and $\alpha$ drop out and we are left with
\begin{align}
\left(\frac{k_B \Delta T_\mathrm{min}}{E_r}\right)\,\left(\frac{\Delta N_\mathrm{tot,min}^2}{N}\right) = \frac{1}{NC\epsilon}. 
\end{align}
This relation implies that for a given minimal integration time, the energy deposited in the system bounds the atom number imprecision and vice versa. The exact value for the minimal integration time and thus the time resolution of the measurement can be chosen by changing the effective scattering rate, i.e. the atom-cavity detuning $\Delta_{ca}$ in our case.
Our observed dependence of the imprecision quantified by the Allan deviation with integration time $\tau$ is shown in Fig.~\ref{fig:1}e, where we find a decrease of the imprecision with increasing integration time for small $\tau<1\,$ms and then an increase for larger $\tau$. We note, however, that there the increase in imprecision is dominated by the dynamically changing mean atom number, which leads to a scaling of the Allan variance with $\tau^2$. Directly comparing the two cases requires subtracting the known time dynamics of the mean atom number, only keeping the stochastic contribution ($\propto \tau$) discussed here. While these two effects can always be separated in post-processing even for an unknown dynamical process by subtracting the mean of all traces, this is generally not possible e.g. for applying feedback to a system with unknown atom number dynamics.

\end{document}